\documentclass[aps,pra,superscriptaddress,twocolumn,longbibliography]{revtex4-1}
\usepackage{bbm}
\usepackage{mathrsfs}
\usepackage{amsmath}
\usepackage{amsfonts}
\usepackage[colorlinks=true,linkcolor=blue,urlcolor=blue,citecolor=blue,anchorcolor=blue]{hyperref}
\usepackage{graphicx,epstopdf}
\usepackage{subfigure}
\usepackage{epsfig}
\usepackage{dcolumn}
\usepackage{bm}
\usepackage{color}
\usepackage{natbib}
\usepackage{amssymb}
\usepackage{xcolor}
\usepackage{braket}
\usepackage{ulem}
\usepackage{float}
\usepackage{braket}
\usepackage{changes}
\usepackage{lipsum}

\begin{document}

\title{Tricritical phenomena induced by higher order Fermi surface nesting in self-organized Fermi gas}
\author{Yilun Xu}
\affiliation{State Key Laboratory for Mesoscopic Physics, School of Physics, Frontiers Science Center for Nano-optoelectronics, Peking University, Beijing 100871, China}
\affiliation{Beijing Academy of Quantum Information Sciences, Beijing 100193, China}
\author{Feng-Xiao Sun}
\email{sunfengxiao@pku.edu.cn}
\affiliation{State Key Laboratory for Mesoscopic Physics, School of Physics, Frontiers Science Center for Nano-optoelectronics, Peking University, Beijing 100871, China}
\author{Qiongyi He}
\affiliation{State Key Laboratory for Mesoscopic Physics, School of Physics, Frontiers Science Center for Nano-optoelectronics, Peking University, Beijing 100871, China}
\affiliation{Collaborative Innovation Center of Extreme Optics, Shanxi University, Taiyuan 030006, China}
\affiliation{Peking University Yangtze Delta Institute of Optoelectronics, Nantong 226010, China}
\affiliation{Hefei National Laboratory, Hefei 230088, China}

\begin{abstract}
	Cold atom systems in optical lattices have long been recognized as an ideal platform for bridging condense matter physics and quantum optics. Here, we investigate the 1D fermionic superradiance in an optical lattice, and uncover novel tricritical phenomena and multistability in finite-temperature cases. As a starting point, we compare the 1D and 2D Fermi gases in zero-temperature limit. It turns out that the tricritical point originates from the higher-order Fermi surface nesting (FSN) and the infrared divergence in 1D systems, which is absent in 2D cases. For dissipative cavities, we obtain the stable phase diagram and observe a hysteresis-type evolution under quench dynamics. When extending to finite-temperature cases, we derive analytical expressions for the 2nd- and 4th-order Landau coefficients, matching the numerical phase diagrams well, and reveal two different trcritical behaviors, named as quantum- and classic-type tricritical points, respectively. In addition, we construct the dependence between the susceptibility and the temperature, giving a nontrivial scaling law $\lim_{\Delta T\rightarrow0}\Delta B_c\sim \Delta T^\nu$ with $\nu>1$. This work provides a new approach to understanding tricritical phenomena, multistability and scaling rate of self-organized cold atoms.
	
	
\end{abstract} 

\maketitle
\section{Introduction.}

With significant advances in laser cooling and trapping techniques~\cite{PhysRevLett.75.37,RN106,RN104,RN105}, a rich variety of condensed matter phenomena have been experimentally simulated in cold atom systems with optical lattices~\cite{RevModPhys.85.553,PhysRevLett.107.277202}, such as the topological phase transitions~\cite{RN108,RN109} and the superfluid phase transition in strongly correlated systems~\cite{RN105,PhysRevLett.92.130403,RN107,annurev}. Different from the conventional avenues to realize long-range interaction through dipolar moment of atoms~\cite{Lahaye_2009}, the cavity-mediated atomic scattering provides infinitely long-range coupling of the quantum gases~\cite{PhysRevA.70.013414,PhysRevLett.89.253003,PhysRevLett.98.203008}. Such cavity-mediated long-range order in atomic ensemble will provide us brand new schemes to simulate the condensed matter and many-body physics, such as quantum frustration~\cite{RN89}, spin liquid~\cite{8qx2-xxh2} and fermionic pairing~\cite{RN90,PhysRevLett.127.177002} and so on. 

Among them, the presence of non-diagonal long-range order indicates non-zero superradiant order parameter in Bose-Einstein condensates (BECs), which was first experimentally realized in 2010~\cite{RN103,PhysRevLett.105.043001,PhysRevLett.104.130401}. It simulated the phase transition of Dicke model by harnessing the momentum states of condensed atoms. As an extended meaning orginated from the spontaneous radiation of atoms~\cite{PhysRev.93.99,GROSS1982301}, the superradiance here refers to the macroscopic number of photons condensing in the cavity. Whereas superradiance in the Dicke model has been extensively studied~\cite{PhysRevLett.107.140402,PhysRevLett.112.173601,PhysRevLett.120.183603,PhysRevLett.122.193201,PhysRevLett.122.193605,PhysRevLett.124.073602,PhysRevLett.128.163601,PhysRevLett.133.233604}, similar phenomena for Fermi gases in cavities --- initially predicted earlier~\cite{PhysRevLett.86.4199,PhysRevLett.86.4203} --- were not observed experimentally until 2021~\cite{doi:10.1126/science.abd4385}. Due to the distinct statistic properties of fermions, a much richer landscape of many-body phases is expected, including the emergence of liquid-gas phase transitions, different scaling behaviors, nonequilibrium dynamic behaviors, and topology ~\cite{PhysRevLett.112.143002,PhysRevLett.112.143003,PhysRevLett.112.143004,PhysRevLett.131.243401,PhysRevA.82.043612,PhysRevLett.112.086401,PhysRevLett.115.045303}. Although the theoretical framework for superradiant phase transitions in optical lattices is well established, open questions remain concerning the self-organization of ultracold atoms, dynamic behaviors, multistabilities and finite-temperature properties.

In this Letter, we concentrate on the fermionic superradiant phenomenon in strong pumping limit, where the scattering of Fermi atoms is confined to the perpendicular plane. We first calculate the second- and fourth-order perturbative contributions in zero-temperature limit analytically, corresponding to two- and four-photon effects, respectively. By comparing the results of 1D and 2D fermi gas, we find that the infrared divergence in 1D systems will result in the tricritical phenomenon, which vanishes in 2D systems as the infrared divergence is absent. With the aid of the analytical results, different scaling behaviors of the critical boundary and the tricritical point are revealed. For a dissipative cavity, a multistable phase diagram is obtained and it's verified by the numerical simulation through master equation, where a hysteresis-type evolution is demonstrated. Additionally, we discuss finite-temperature scenarios, where the 2nd- and 4th-order coefficients in Landau theory are analytically obtained. It reveals that the tricritical point will extend into a tricritical curve with increasing temperature, giving us novel finite-temperature phase diagrams where both quantum- and classical-type tricritical points emerge. What's more, in zero temperature limit, we prove the critical pumping strength is independent of the temperature in 1st order, implying
	$\lim_{\Delta T\rightarrow0}\Delta B_c\sim \Delta T^\nu$ with the scaling rate $\nu>1$.

\begin{figure}[tb]
	\centering
	\includegraphics[width=0.48\textwidth]{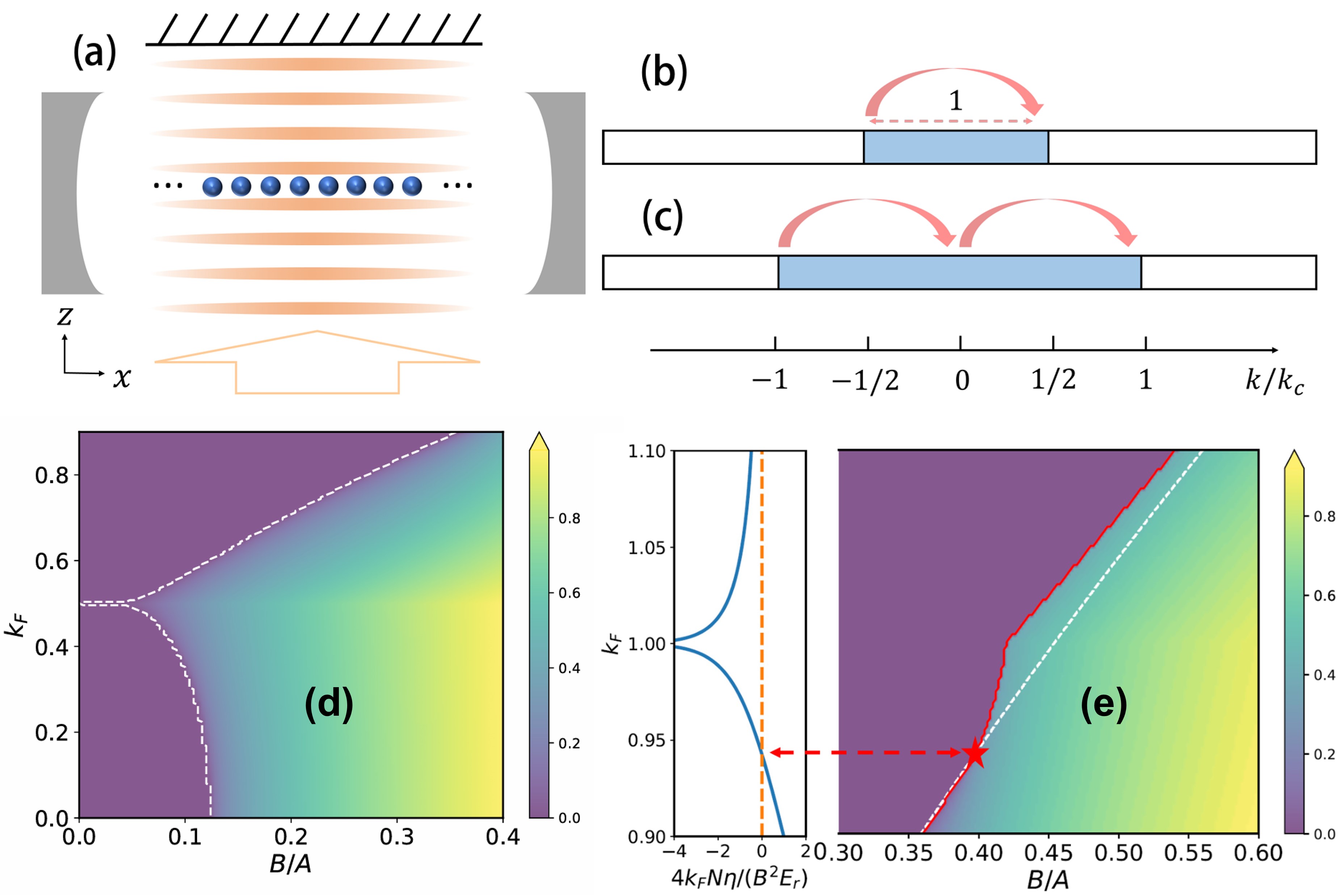}\vspace{-0.1cm}
	\caption{(a) The experimental schematic. The contribution of (b) the first order FSN with single-photon processes, and (c) the second order FSN where two-photon processes are considered. The blue and blank parts stand for the filled and unfilled regions in momentum space. The pink arrow represents the momentum transition after single photon scattering. (d)(e) The phase diagrams for the 1D fermion gas. The white dashed curves represent the disappearance of the susceptibility $\chi$, the red solid curve is the critical boundary, and the blue solid line on left panel is the renormalized fourth-order coefficients.We set $A=1$ here and in the following text.}\label{schematic}
\end{figure}



\section{Self-organized Fermi gas in strong pumped optical lattice}

We consider a macroscopic number of spinless fermions placed in the center of a cavity as shown in Fig.~\ref{schematic}(a), which is an experimental scenario similar to Ref.~\cite{PhysRevLett.112.143002,PhysRevLett.112.143004}. In order to stimulate the coupling between the atoms and the single transverse cavity mode in $x$ direction, the system is pumped by a strong standing-wave laser along $z$ direction. We denote the optical mode as $a$, the atomic ground and excited states as $\ket{g}$ and $\ket{e}$, respectively. The system Hamiltonian $H$ is expressed as $H=H_0+H_I$, with
\begin{align}
	H_0&=\omega a^\dagger a+\omega_{e}\ket{e}\bra{e},\\
	H_I&=-\Omega\sigma^+e^{-i\omega_{p}t}\cos(\vec{k}_{p}\cdot\vec{r})-g\sigma^+a\cos(\vec{k}_c\cdot\vec{r})+H.c..
\end{align}
Here, we set $\hbar=1$. $g$ is the coupling strength between fermions and cavity mode $a$ with wave vector $\vec{k}_c$ (in $x$ direction) and eigenfrequencies $\omega$. $\Omega$ represents the Rabi frequency driven by the pump laser with wave vector $\vec{k}_{p}$ (in $z$ direction) and frequencies $\omega_{p}$. $\sigma^+$ stands for the Pauli raising operator $\ket{e}\bra{g}$. For simplicity, the ground-state energy of $\ket{g}$ is set to zero and the transition frequency between the ground and excited states to $\omega_{e}$.

Defining $\tilde{\omega}\equiv\omega-\omega_{p}$ and $\Delta\equiv\omega_{e}-\omega_{p}$ as the cavity and atom detunings against the pump laser, respectively, we can apply the adiabatic elimination for the excited state $\ket{e}$ and obtain the effective Hamiltonian as
\begin{align}\label{H_tot0}
	H_{tot}&=\tilde{\omega}a^\dagger a+\sum_{\vec{k}}\dfrac{\vec{k}^2}{2m}c_{\vec{k}}^\dagger c_{\vec{k}}-\dfrac{g\Omega}{4\Delta\sqrt{N}}(a+a^\dagger)\sum_{\vec{k},s}c_{\vec{k}}^\dagger c_{\vec{k}+s\vec{k}_c}.
\end{align}
Here, the cavity-atom coupling strength $g$ has been changed to $g/\sqrt{N}$ for a technique reason, which is explained in the Supplemental Material (SM)~\cite{supplementary}. In addition, the strong pump limit requires $\Omega^2\gg\Delta E_r$ with $E_r={k_c^2}/(2m)$ being the recoil energy.

In thermodynamic limit, where the total atom number $N\rightarrow\infty$, the operator of the optical field $a$ can be substituted by its expectation value $\left<a\right>\equiv\psi$ based on the coherent state approximation. Hence, the Hamiltonian~(\ref{H_tot0}) is viewed as a single particle fermionic Hamiltonian $H_{tot}(\psi)$ dependent on the order parameter $\psi$. Meanwhile, the system can be viewed as the canonical ensemble by ignoring the atom number leakage, in our scenario. The chemical potential $\mu$ is decided by the unchanged atom number as $N=\sum_{\vec{k}}n(\vec{k})$, where $n(\vec{k})=[e^{\beta(\epsilon_{\vec{k}}-\mu)}+1]^{-1}$ and $\epsilon_{\vec{k}}$ is the eigenvalue of the Hamiltonian $H_{tot}(\psi)$. Obviously, the chemical potential $\mu$ will be changed with $T$ and $\psi$.

\begin{figure}[tb]
	\centering
	\includegraphics[width=0.48\textwidth]{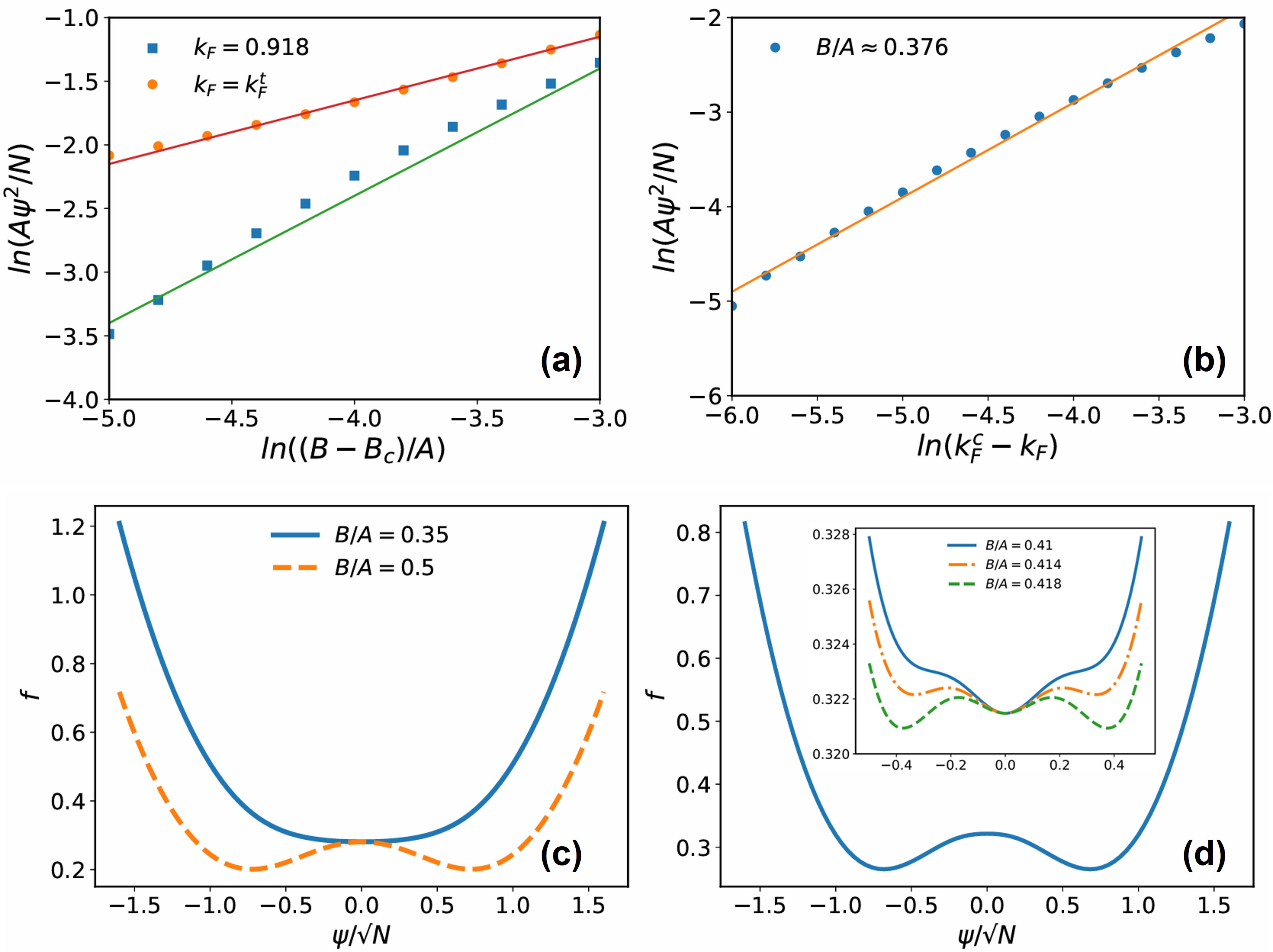}\vspace{-0.1cm}
	\hspace{-0.3cm}
	\caption{(a)The critical scaling behavior for $k_F=0.918$(orange dots) and $k_F=k_F^t=0.9425$ (blue square dots), respectively. The red and green solid lines stand for the linear fitting for the ramping rate $1/2$ and $1$. (b)The critical scaling for fixed $B/A\approx0.376$, and the ramping rate is around $1$. The average free energy $f$ curve changes against the order parameter $\psi/\sqrt{N}$ with $k_F=0.918$, and $B/A=0.35,0.5$ in subfigure (c); $k_F=0.982$, $B/A=0.5$ in subfigure (d), and $B/A=0.41,0.414,0.418$ in the inset. }
	\label{free_energy_fig}
\end{figure}

\section{Zero-temperature Fermionic superradiance.}
Recall that the free energy $F$ is equal to the ground-state energy of the system in zero-temperature limit~\cite{independentmu}, which means that the expression of $F$ is independent of the chemical potential $\mu$ as $T\to0$, i.e., the part of the contribution $\dfrac{\partial F}{\partial \mu}\dfrac{\partial \mu}{\partial (\psi^2)}=0$. 
Then we can expand the free energy of the system till the fourth-order term of the optical field as
\begin{align}\label{free energy_0T}
	F=\tilde{\omega}\left|\psi\right|^2+\chi(\psi+\psi^*)^2+\eta(\psi+\psi^*)^4+\dots,
\end{align}
where the Landau's coefficients $\chi$ and $\eta$ are available from the second and fourth perturbation contributions, respectively. By minimizing the free energy~(\ref{free energy_0T}), we obtain that $\psi=0$ if $\tilde{\omega}+4\chi>0$, and $\psi^2=-({\tilde{\omega}+4\chi})/({32\eta})$ if $\tilde{\omega}+4\chi<0$ with the assumption of $\eta>0$. 

According to the expression of $\chi$[see in the appendix], the critical point is achieved as
\begin{align}
	k_F=2(B/A)\ln\left|(k_F+1/2)/(-k_F+1/2)\right|,
\end{align}
which means that the critical pumping strength $B/A$ is only dependent on the filling factor $k_F$. Here $A\equiv\tilde{\omega}/E_r$ and $B\equiv UV/E_r^2$ are defined as dimensionless parameters with effective coupling strengths $V\equiv{\Omega^2}/({4\Delta})$, $U\equiv{g^2}/({4\Delta})$. And a remarkable divergence can be observed in Fig.~\ref{schematic}(d) near the boundary of the first Brillouin zone, where superradiance can be observed even with a tiny coupling strength $B$. This phenomenon is attributed to the first-order Fermi surface nesting (FSN) shown in Fig.~\ref{schematic}(b), indicating an almost zero energy cost for fermionic scattering by a single cavity photon across opposite sides of the Fermi surface. In the low occupation limit $k_F\to0$, we have $B/A=1/8$ at the critical point, which matches the critical boundary for boson gases.

Remarkably, we find $\eta$ manifests different behaviors in 1D and 2D gases, respectively. In 1D systems, we have
\begin{align}
	\eta^{1D}\propto2\ln\left|\frac{k_F+1/2}{-k_F+1/2}\right|-\ln\left|\frac{k_F+1}{-k_F+1}\right|+\frac{k_F}{2(1/4-k_F^2)^2},
\end{align}
exhibiting infrared-divergent behaviors in both $k_F\sim1/2$ and $k_F\sim1$ with opposite signs. When it comes to 2D system,
\begin{align}
	\eta^{2D}>0
\end{align}
will be valid [see in the Appendix]. Recall that the tricritical point emerges at the intersection of $\tilde{\omega}+4\chi=0$ and $\eta=0$. Hence, the tricritical point exists in 1D systems, but is absent in 2D systems. We demonstrate the phase diagram around the tricritical point as in Fig.~\ref{schematic}(e). The white dashed line represents the boundary $\tilde{\omega}+4\chi=0$. And the red solid line stands for the critical boundary between normal phase (NP) and superradiant phase (SRP) deviating from the white dashed line at the tricritical point (red star) with $\eta=0$. The critical line below and above the tricritical point manifests different behaviors of both the order parameter and the free energy behaviors.


\begin{figure}[tb]
	\centering
	\hspace{-0.3cm}
	\includegraphics[width=0.5\textwidth]{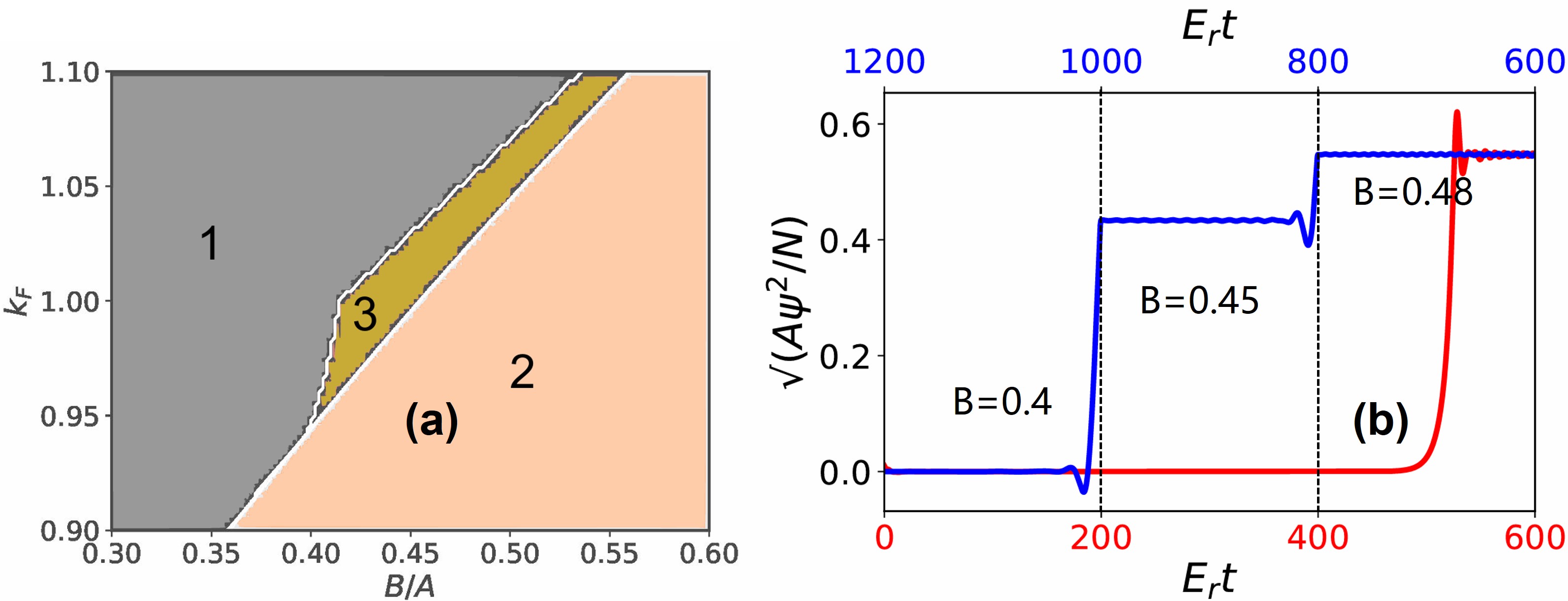}
	\caption{(a) The stable-state phase diagram in the limit $\kappa\to 0$. The grey, orange and brown regions support $1$, $2$ and $3$ stable states, respectively, labeled by the digits. (b) The quench dynamic evolution of the order parameter with the forward and backward three-step pulse, marked by red and blue lines. The filling factor is fixed as $k_F=0.98$.}\label{pd}
\end{figure}

In the zero-temperature limit, the tricritical point is located at $k_F^t(T=0)\approx0.9425$, $B^t(T=0)/A\approx0.395$, where the scaling law is absolutely different from the normal critical boundary [see in the Appendix]. It turns out the scaling rate gives
\begin{align}
	\sqrt{A}\psi/\sqrt{N}&\sim\left|(B-B_c)/A\right|^{\nu}\\
	\sqrt{A}\psi/\sqrt{N}&\sim\left|k_F^c-k_F\right|^{1/2},
\end{align}
with $\nu=1/2$ in the normal critical boundary and $\nu=1/4$ at tricritical point, which are shown in Fig.~\ref{free_energy_fig}(a)(b), respectively. Here $B_c$ and $k_F^c$ represent the critical (tricritical) point for the pumping strength and filling factor separating the NP and SRP.

Then, we consider the average free energy $f\equiv F/N$ changes against the order parameter $\psi/\sqrt{N}$ at different parametric points in the phase diagram. In Fig.~\ref{free_energy_fig}(c), we observe that one global minimum point of the free energy will be splitted into two degenerate global minima after the second order phase transition. And in Fig.~\ref{free_energy_fig}(d), the free energy curves around the first order phase transition manifest richer phenomena. As the ratio $B/A$ is far from the phase boundary, the system has only one global minimum. With the ratio $B/A$ increasing, the second-order coefficient becomes smaller, and the negative contribution from the fourth-order term will lead to two local minima in both sides of the original point. Although the global minimum is still at the original point $\psi=0$, these two local minimum points also support two stable states in dissipative cases, resulting in multistability which will be mentioned next. With $B/A$ increasing, the two local minimums eventually become two degenerate global minimum points, and the original point will be the local stable state. As the parameter enters the right side of the white dashed line in the phase diagram, where the second-order coefficient is negative, the NP state $\psi=0$ becomes unstable and only two superradiant states exist in this case.


	Note that tricritical phenomena in 2D systems have been studied with weak pumping assumption~\cite{PhysRevLett.112.143002}, where the 2nd order perturbative calculation was enough to predict the tricritical behavior. But here we focus on strong pumping limit, which requests us to analyze higher order perturbative effect, i.e., the 2nd-order FSN. Instead of single-photon scattering described by the 1st-order FSN, now the Fermi gas suffers from strong two-photon scattering near the second Brillouin boundary [Fig.~\ref{schematic}(c)]. A nontrivial finding is that the scattering effect also influenced by the dimension of Fermi gas. In 1D system, the 2nd-order FSN gives the infrared divergence, leading to negative $\eta$ as $k_F\gtrsim1$. While such infrared divergence will disappear due to an additional integral dimension in 2D systems, thus the tricritical point will be absent.




\section{Multistability and hysteresis dynamic in dissipative cavity.} 
	To facilitate comparison, we apply the numerical iterative method to solve the self-consistent equation [see in the appendix], the results of which can be viewed as the stable states of the self-consistent equation in dissipative cavity with $\kappa\to0$. The stable-state phase diagram is illustrated in Fig.~\ref{pd}(a), where the digits indicate the number of stable states supported in the corresponding regions. Surprisingly, in the middle region beyond the tricritical point, marked by $3$, both NP and SRP phases can be stabilized, where the final stable state depends on the initial state of the system. In another aspect, all the local minimums of the free energy should be carefully considered, whose number coincides with the number of stable states obtained by iterative methods [See details in the Appendix]. And such phenomena are also well explained by the free energy plots in Fig.~\ref{free_energy_fig}(c)(d) discussed above.
	
	To further verify the multistability and observe its dynamic evolution, we carry out a three-step quench pulse $B(t)=0.4\to045\to0.48$ with fixed interval $E_r\Delta t=200$ in forward evolution (red line) and $B(t)=0.48\to045\to0.4$ with the same interval in backward evolution (blue line), shown in Fig.~\ref{pd}(b). In each pulse interval, the order parameter will finally reach the stable state featured by a parallel and smooth segment. Obviously, for $B=0.45$ located in region 3 of Fig.~\ref{pd}(a), the forward and backward pulses lead the system to different stable states, giving NP and SRP stable states respectively.



\begin{figure}[tb]
	\centering
	\includegraphics[width=0.48\textwidth]{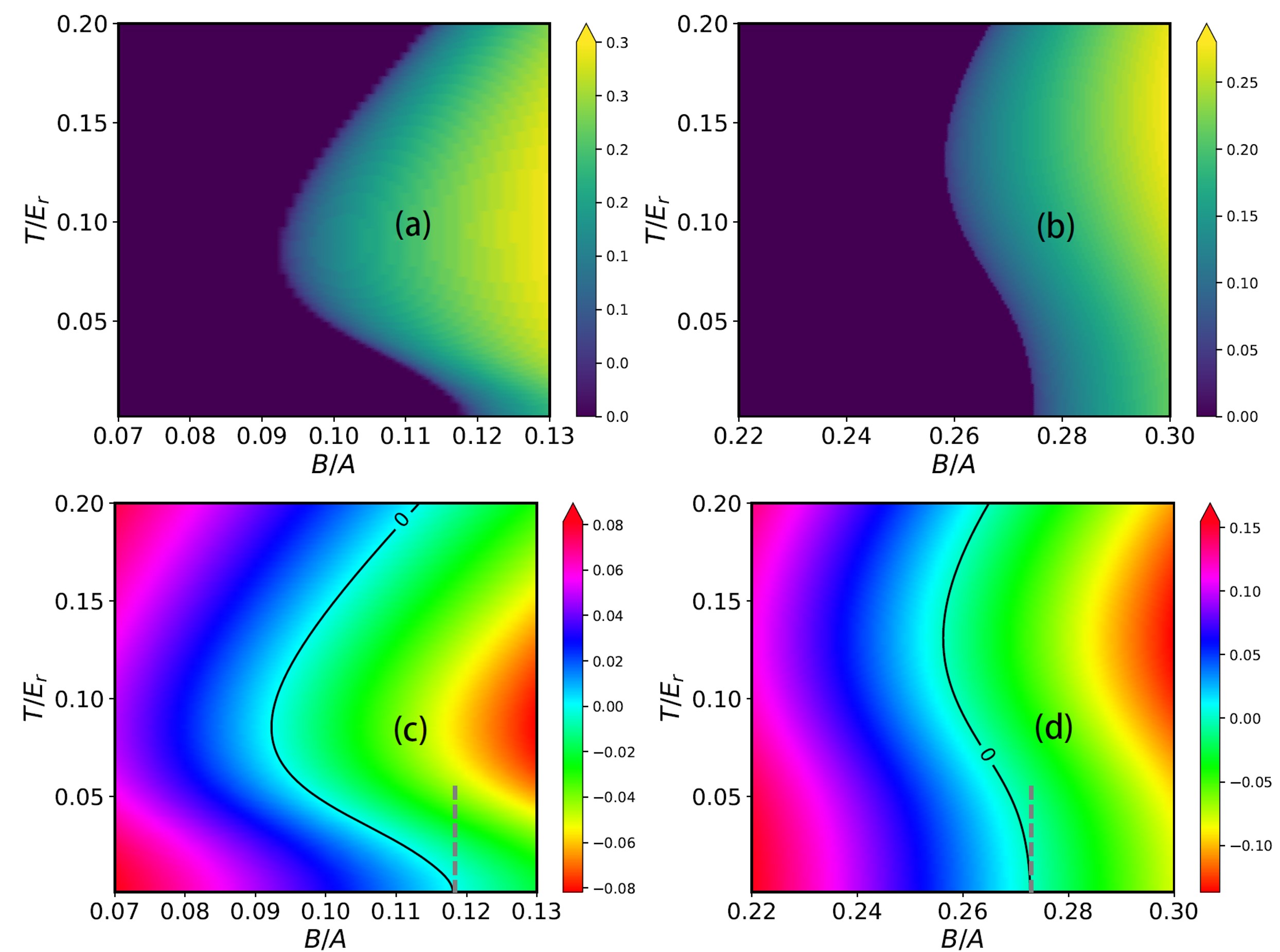}
	\caption{The numerical equilibrium phase diagram at finite temperatures with fixed filling factors $k_F=0.2$ (a) and $k_F=0.8$ (b), and the  corresponding analytical colormaps of $k_F(\tilde{\omega}+4\chi)/E_r$ are shown in (c) and (d), respectively. The grey dashed lines are the tangents lines at $T=0$.}\label{finite_T_pd}
\end{figure}


\section{Finite-temperature phase diagram and scaling behavior against temperature.} For a finite-temperature scenario, it is easily checked that $\chi<0$ at arbitrary temperature $T$, implying that superradiance can potentially be observed with sufficiently strong atom-cavity coupling and pumping laser intensity, even at high temperatures. 
And a more interesting question is how the critical or tricritical point evolves with increasing temperature. We present the phase diagram on $T$-$B$ parameter plane in Fig.~\ref{finite_T_pd}, with the filling factor $k_F$ fixed in each slice. Although there is no strict Fermi surface in finite-temperature cases, we still use $k_F$ to represent the fermionic filling number through the relation $N=\sum_{\vec{k}}n(\vec{k})=\sum_{\left|\vec{k}\right|<k_F}$. 

As shown in Fig.~\ref{finite_T_pd} the changing tendency from $T\rightarrow0$ to $T\neq0$ is smooth, indicating the continuity of susceptibility $\chi$. When it comes to the first order dependence between $\chi$ and $T$, we give a detailed derivative to prove that $\dfrac{\partial\chi}{\partial T}\Bigg|_{\psi=0,T\rightarrow0}=0$, and $\dfrac{dB_c}{dT}\Bigg|_{T_c\rightarrow0}=0$ [see in the Appendix]. This tells us a nontrivial fact that the critical boundary at finite temperature won't deviate from the zero temperature one, at least in terms of the first order of temperature. Thus, we obtain the scaling rate $\lim_{\Delta T\rightarrow0}\Delta B_c\sim \Delta T^\nu$ with $\nu>1$, where $\Delta B_c\equiv B_c({T=\Delta T})-B_c({T=0})$.
	
	Moreover, the numerical critical points in Fig.~\ref{finite_T_pd}(a)(b) are consistent with the analytical zero contour lines of the second-order coefficient $\tilde{\omega}+4\chi$ in Fig.~\ref{finite_T_pd}(c)(d). It can be seen that the tangent lines of the critical boundaries around $T_c\rightarrow0$ points are parallel to the $T$ axis, making our analytical conclusion convincing. Furthermore, it is found that the minimum coupling strength $B$ to observe superradiance does not occur at zero temperature. Instead, there exists an optimal temperature where the superradiant phase transition is most accessible.

\begin{figure}[tb]
	\hspace{-0.8cm}
	\includegraphics[width=0.48\textwidth]{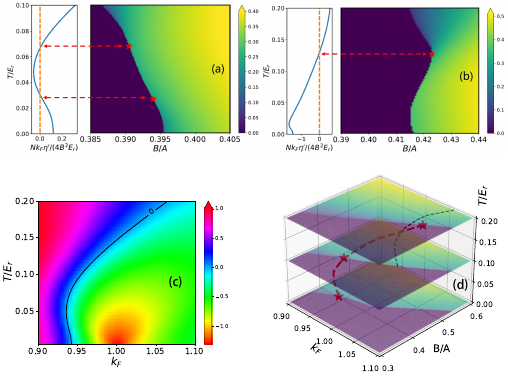}
	\caption{The phase diagram at finite temperatures with fixed filling factors $k_F=0.938$ (a), $k_F=0.982$ (b). The colorbar represents the order parameter, and the red stars stand for the tricritical points, corresponding to the zero points of $\eta'$ (blue line on left panel). (c) The analytical fourth order coefficients $Nk_F\eta'/(4B^2E_r)$ colormap with black line referring to the zero contour line. (d) The 3D phase diagram where three slices with fixed temperature $T=0,0.1,0.2$ are displayed. The red dashed curve is connected through all the tricritical points in every slices between $T=0$ and $T=0.2$, whose projection is shown with black dashed line in the $T$-$k_F$ plane.}
	\label{3D_tricritical_curve}
\end{figure}


\section{Categories of tricritical points in finite temperature}

When focusing on tricritical phenomena, Fig.~\ref{3D_tricritical_curve}(a) shows two tricritical points in the $k_F=0.938$ slice [see in the Appendix]. With increasing temperature, the $2nd \to 1st$ and $1st \to 2nd$ type tricritical points can be viewed as the quantum- and classical-type tricritical points, respectively, according to whether the transition temperature is low or high. If the temperature is low, the Fermi surface will be roughly preserved with the momentum distribution just slightly enlarged. Thus, we observe a similar $2nd \to 1st$ phase transition in Fig.~\ref{3D_tricritical_curve}(a) as that around $k_F=0.9425$ in Fig.~\ref{pd}(b), which is dominated by quantum effects from the 2nd order FSN. In contrast, at higher temperature, the Fermi distribution broadens and becomes more like a classical one, the 2nd order FSN effect will be submerged in the whole contribution of all the atoms.
Meanwhile, for a fixed filling factor $k_F=0.982$ located in the first-order phase transition region at zero temperature [Fig.~\ref{pd}(b)], Fig.~\ref{3D_tricritical_curve}(b) shows that only one tricritical point emerges, corresponding to the classical type. 

In order for a more intuitional understanding on the finite-temperature system, we derive the analytic expression for the 4th order coefficients of the order parameter $\psi$, defined as $\eta'\equiv\dfrac{\partial^2F}{2\partial(\psi^2)^2}\Bigg|_{\psi=0}$ [see in the Appendix], and the corresponding colormap is shown in  Fig.~\ref{3D_tricritical_curve}(c). The $\eta'=0$ contour line is marked in black, which corresponds to tricritical points at appropriate pumping strength $B$. We find the curve first evolves toward smaller $k_F$, reaches a minimum at $\min\left\{k_F^t\right\}=k_F^t(T\approx0.048)\approx0.9358$, and then turns toward larger $k_F$. We also connect all tricritical points in every slice numerically with fixed $T$ and obtain a 3D tricritical curve in Fig.~\ref{3D_tricritical_curve}(d). By projecting in $T$-$k_F$ plane, we find it closely matches the $\eta'=0$ contour line in Fig.~\ref{3D_tricritical_curve}(c). In this aspect, the tricritical points in Fig.~\ref{3D_tricritical_curve}(a)(b) correspond to the intersections between the tricritical curve and the $k_F=0.938,0.982$ slices.


Hence, all possible cases can be divided into three categories according to different filling factors $k_F$. When $\min\left\{k_F^t\right\}<k_F<k_F^t(T=0)$, both the quantum- and classical-type tricritical phenomena are observed with increasing temperature $T$. And these two types of tricritical points coalesce in the $\min\left\{k_F^t\right\}$ slice. Thus, if $k_F<\min\left\{k_F^t\right\}$, only the 2nd order phase transition can be observed without a tricritical phenomenon. And when $k_F>k_F^t(T=0)$, only the classical-type tricritical point can be revealed at high temperature. 


\section{Conclusion}
In this paper, we study the superradiant phase transition induced by a strongly confined spinless Fermi gas in an optical lattice. In this scenario, we analytically derive the single- and two-photon scattering contributions for 1D and 2D configurations on the confined plane perpendicular to the pumping laser. We find that, in 1D cases, the 1st- and 2nd-order FSN contributions exhibit infrared divergences around $k_F=1/2$ and $k_F=1$ with opposite signs, such that a tricritical point emerges at the intersection between curves $\chi=0$ and $\eta=0$. While in 2D cases, the additional integral dimension renders the fourth-order coefficient convergent, resulting in $\eta>0$. Therefore, the tricritical point is only restricted to 1D systems. We also present the stable phase diagram for dissipative cavities and observe the hysteresis-type dynamics, confirming the existence of multistability. In addition, at finite temperature, we analytically derive the 2nd- and 4th-order coefficients and find excellent agreement with numerical phase diagrams. Remarkably, the tricritical points extend into a tricritical curve as $T$ varies. Through both quantitative and qualitative analyses, quantum- and classical-type tricritical points are observed and explained. By examining the temperature dependence of the critical boundary in the zero-temperature limit, we rigorously obtain a scaling rate $\nu>1$, which is first explored in fermionic superradiant systems to our knowledge. Furthermore, we find that there exists an optimal temperature, instead of zero temperature, at which the superradiant phase transition is most readily realized, facilitating the experimental observation of such critical phenomena. Overall, our work provides new insights into quantum simulation in optical lattice systems, broadens the understanding of atomic self-organization, reveals rich nonequilibrium many-body dynamics, and deepens the knowledge of finite-temperature fermionic superradiance, thereby advancing the exploration and manipulation of novel quantum materials mediated by cavity~\cite{PhysRevLett.84.4068,PhysRevLett.125.053602,PhysRevLett.127.177002,PhysRevLett.130.083603,PhysRevA.87.023831,PhysRevB.103.075131}.

\begin{acknowledgments}
	We thank Han Pu from Rice University for his instructive suggestions. This work is supported by the National Natural Science Foundation of China (Grant Nos. 12125402, 12474256, 12534016), Beijing Natural Science Foundation (Grant No. Z240007), and the Innovation Program for Quantum Science and Technology (Nos. 2024ZD0302401 and 2021ZD0301500).
\end{acknowledgments}

\onecolumngrid

\appendix

\section{derive the effective Hamiltonian of the system}

By means of the unitary transformation $U(t)=\exp[i(\ket{e}\bra{e}+a^\dagger a)\omega_{p}t]$, the interaction Hamiltonian in the rotation frame can be written as 
\begin{align}
	\tilde{H}_I&=\tilde{\omega}a^\dagger a+\Delta\ket{e}\bra{e}-[\Omega\sigma^+\cos(\vec{k}_{p}\cdot\vec{r})+g\sigma^+a\cos(\vec{k}_c\cdot\vec{r})+H.c.],
\end{align}
where $\tilde{\omega}\equiv\omega-\omega_{p}$ and $\Delta\equiv\omega_{e}-\omega_{p}$ stand for cavity and atom detunings against the pump laser, respectively. Considering the dynamic evolution of the eigenbasis of the fermion inner states $\left\{\ket{g},\ket{e}\right\}$, we can write the equations according to the Schr\"{o}dinger equations driven by the Hamiltonian $\tilde{H}_I$,
\begin{align}
	&i\dfrac{d}{dt}\ket{g}=-\Omega \cos(\vec{k}_{p}\cdot\vec{r})\ket{e}-ga\cos(\vec{k}_c\cdot\vec{r})\ket{e},\\
	&i\dfrac{d}{dt}\ket{e}=\Delta\ket{e}-\Omega\cos(\vec{k}_{p}\cdot\vec{r})\ket{g}-g\cos(\vec{k}_c\cdot\vec{r})a^\dagger\ket{g}.
\end{align}
Assuming that the detuning $\Delta$ is large enough, we can apply adiabatic elimination for the excited states $\ket{e}$, and the effective dynamic evolution for the ground state can be obtained approximately as
\begin{align}
	i\dfrac{d}{dt}\ket{g}=&-\dfrac{1}{\Delta}[\Omega^2\cos^2(\vec{k}_{p}\cdot\vec{r})+g^2a^\dagger a\cos^2(\vec{k}_c\cdot\vec{r})+\Omega g(a+a^\dagger)\cos(\vec{k}_{p}\cdot\vec{r})\cos(\vec{k}_c\cdot\vec{r})]\ket{g}.
\end{align}

Then, the effective Hamiltonian after the adiabatic elimination is expressed as follows:
\begin{align}
	\tilde{H}_I\approx\tilde{H}_I^{ad}=&[\tilde{\omega}-\dfrac{g^2}{\Delta}\cos^2(\vec{k}_c\cdot\vec{r})]a^\dagger a-\dfrac{g\Omega}{\Delta}(a+a^\dagger)\cos(\vec{k}_c\cdot\vec{r})\cos(\vec{k}_{p}\cdot\vec{r})-\dfrac{\Omega^2}{\Delta}\cos^2(\vec{k}_{p}\cdot\vec{r}).
\end{align}
Taking the kinetic energy of fermions into account, the total Hamiltonian can be written as
\begin{align}\label{H_tot}
	H_{tot}\approx&\tilde{\omega}a^\dagger a+\sum_{\vec{k}}\dfrac{\vec{k}^2}{2m}c_{\vec{k}}^\dagger c_{\vec{k}}-\dfrac{g^2}{4\Delta}a^\dagger a\sum_{\vec{k},s=\pm1}c_{\vec{k}}^\dagger c_{\vec{k}+2s\vec{k}_c}-\dfrac{g\Omega}{4\Delta}(a+a^\dagger)\sum_{\vec{k},s,s'=\pm1}c_{\vec{k}}^\dagger c_{\vec{k}+s\vec{k}_c+s'\vec{k}_{p}}\notag\\
	&-\dfrac{\Omega^2}{4\Delta}\sum_{\vec{k},s=\pm1}c_{\vec{k}}^\dagger c_{\vec{k}+2s\vec{k}_{p}}.
\end{align}
Here, the sign of "$\approx$" originates from the assumption that $\tilde{\omega}\approx\tilde{\omega}-\dfrac{g^2}{2\Delta}$ by consider a rather small coupling strength $g$ reasonably. In this work, we concentrate on the case where the pump laser is strong enough, so that the Fermi gas is highly localized in $z$ direction. Since the wave vector of the cavity mode is perpendicular to the pump laser, the fermi gas can be viewed as a 2D ensemble in $x$-$y$ plane, and the dispersive in pump direction $z$ is flat. This can be given with an equivalent description as $\dfrac{\Omega^2}{\Delta E_r}\gg1$, where $E_r=\dfrac{k_c^2}{2m}$ is the recoil energy, and the last term is negligible. As a technique consideration, we change the cavity-atom coupling strength $g$ to $g/\sqrt{N}$, where $\sqrt{N}$ in the denominator is also called a renormalization factor, making the results independent of the macroscopic atom number $N$. And we also assume that $\dfrac{g^2}{4\Delta}\ll\tilde{\omega},E_r$ so that the third term on the right hand side of Eq.~(\ref{H_tot}) can hardly affect the background potential of the Fermi gas. Based on the considerations above, the Hamiltonian can be further simplified as 
\begin{align}
	H_{tot}&=\tilde{\omega}a^\dagger a+\sum_{\vec{k}}\dfrac{\vec{k}^2}{2m}c_{\vec{k}}^\dagger c_{\vec{k}}-\dfrac{g\Omega}{4\Delta\sqrt{N}}(a+a^\dagger)\sum_{\vec{k},s}c_{\vec{k}}^\dagger c_{\vec{k}+s\vec{k}_c}.
\end{align}

In aspect of perturbation theory, we consider the strength coefficient of the 4th order term of optical mode such as $a^4,a^3a^{\dagger},\dots$, which is the lowest order contribution from the third term of Eq.~(\ref{H_tot}). The third term will give the coefficient $\sim E_r(\dfrac{g^2}{\Delta E_r})^2$, while the fourth term will give a much stronger strength $\sim E_r(\dfrac{g\Omega}{\Delta E_r})^4$, If we have $\dfrac{\Omega^2}{\Delta E_r}\gg1$ as mentioned before. And the same conclusion can be obtained if we consider all order contributions from the third term of Eq.~(\ref{H_tot}), indicating the third term can be ignored safely. Therefore, this is a another understanding on why we drop both the third and last term of Eq.~(\ref{H_tot}).

\section{calculation of the 2nd and 4th order pertubative contribution}

\begin{figure}[tb]
	\centering
	\includegraphics[width=1.0\textwidth]{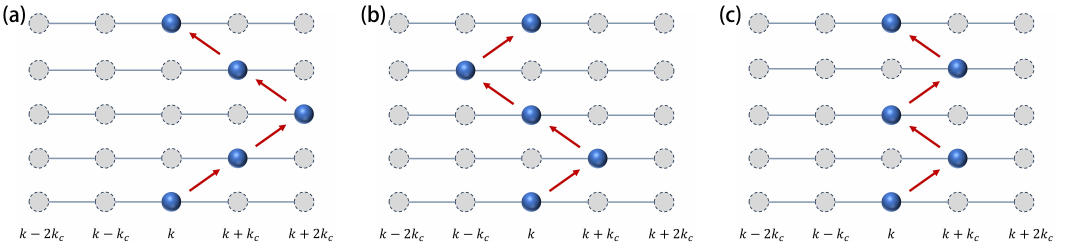}
	\caption{The schematic figure for the fourth order perturbative contribution of $\tilde{\eta}_1,\tilde{\eta}_2,\tilde{\eta}_3$ in (a-c) respectively. We demonstrate five coupled momentum states with blue solid line connected. The red solid arrows instruct the scattering paths for single fermion in fourth order perturbative process.}
	\label{eta_schematic}
\end{figure}


The grant partition function for the fermion part can be expressed as
\begin{align}
	\label{partition function}
	Z_G&=\int\mathcal{D}[\left\{\zeta_{\vec{k}}\right\}]\exp(-S_1)\equiv \exp(-\tilde{S}_1).
\end{align}
Here, we can easily express $\tilde{S}_1=-\ln(Z_G)$ as the counterpart of the grant potential in field theory. And the free energy of the system can be expressed as 
\begin{align}
	\label{free energy}
	F\equiv U-TS=J+N\mu=\tilde{\omega}\left|\psi\right|^2-k_BT\ln Z_G+N\mu,
\end{align}

We are supposed to evaluate the grant potential $\tilde{S}_1$ in Eq.~(\ref{partition function}). Furthermore, we can give a more compact form for $S_1$ as
\begin{align}
	\label{action S1}
	S_1&=\int_{0}^{\beta}d\tau \Big[\sum_{\vec{k}}\zeta_{\vec{k}}^*(\partial_\tau+\dfrac{\vec{k}^2}{2m}-\mu)\zeta_{\vec{k}}-\dfrac{g\Omega}{4\Delta\sqrt{N}}(\psi+\psi^*)\sum_{\vec{k},s}\zeta_{\vec{k}}^*\zeta_{\vec{k}+s\vec{k}_c}\Big]\notag\\
	&=\int_{0}^{\beta}d\tau\left[\vec{\zeta}^\dagger G_0^{-1}\vec{\zeta}+\vec{\zeta}^\dagger V\vec{\zeta}\right].
\end{align}
Here, $[G_0^{-1}]_{\vec{k},\vec{k}'}=(\partial_\tau+\dfrac{\vec{k}^2}{2m}-\mu)\delta_{\vec{k},\vec{k}'}$. The Grassmann vector $\vec{\zeta}$ is composed of all Grassmann fields labeled by different momentum indices $\vec{k}$. And the interaction matrix is represented by $[V]_{\vec{k},\vec{k}'}=-\dfrac{g\Omega}{4\Delta\sqrt{N}}(\psi+\psi^*)\delta_{\vec{k},\vec{k}'\pm\vec{k}_{c}}$. With the aid of the Matsubara Fourier transformation, the integral can be calculated order-by-order after perturbative expansion. We set the Grassmann fermionic field in the Matsubara frequency region as $\vec{\zeta}_n=\dfrac{1}{\beta}\int_0^\beta d\tau\vec{\zeta}(\tau)e^{iv_n\tau}$,
where $v_n=(2n+1)\pi/\beta$ is the Matsubara frequency. While we assume that the boson field is separable with the fermion field, i.e., we can view that $\partial\psi_i/\partial\tau\approx0$, which is also called the long-wavelength limit. So we can get
\begin{align}
	\tilde{S}_1&=-\ln(Z_G)\notag\\
	&=-\ln[det(\beta(G_0^{-1}+V))]\notag\\
	&=-Tr[\ln(\beta(G_0^{-1}+V))]\notag\\
	&\approx-Tr[\ln(\beta G_0^{-1})]+\dfrac{1}{2}Tr[(G_0V)^2]+\dfrac{1}{4}Tr[(G_0V)^4],
\end{align}
where the first equation $Z_G=det(\beta(G_0^{-1}+V))$ is derived from the Gaussian functional integral for the Grassmann field. And the notation of ``Tr'' includes the summation for different Matsubara components and different momentum components. In the last line, only the second and fourth orders of the interaction $V$ are retained here (only the even order terms satisfy the momentum conservation, giving non-zero results). The first term corresponds to the zeroth-order contribution for the free Fermi gas. In the zero-temperature limit, it turns out that the first term cancels the chemical potential $\mu$ in the expression of free energy (\ref{free energy}) as
\begin{align}
	&\lim_{\beta\to\infty}-\beta^{-1}Tr[\ln(\beta G_0^{-1})]+N\mu=\sum_{\vec{k}}(\dfrac{\vec{k}^2}{2m}-\mu)+N\mu=\sum_{\vec{k}}\dfrac{\vec{k}^2}{2m},
\end{align}
giving only the kinetic energy of the fermions which obeys the quadratic dispersion. Then we can expand the free energy of the system till the fourth order term of the optical field as
\begin{align}
	F=\tilde{\omega}\left|\psi\right|^2+\chi(\psi+\psi^*)^2+\eta(\psi+\psi^*)^4.
\end{align}
Obviously, the coefficients $\chi$ and $\eta$ originate from the second and fourth perturbation contributions $\dfrac{1}{2}Tr[(G_0V)^2]$ and $\dfrac{1}{4}Tr[(G_0V)^4]$, respectively. To minimize the free energy above, we obtain that $\psi=0$ if $\tilde{\omega}+4\chi>0$, and $\psi^2=-\dfrac{\tilde{\omega}+4\chi}{32\eta}$ if $\tilde{\omega}+4\chi<0$ and $\eta>0$.

Then, we calculate the second-order contribution as follows:
\begin{align}
	Tr[(G_0V)^2]&=\sum_n\sum_{\vec{k},\vec{k}'}(iv_n-\epsilon_{\vec{k}})^{-1}(iv_n-\epsilon_{\vec{k}'})^{-1}V_{\vec{k}\vec{k}'}V_{\vec{k}'\vec{k}}\notag\\
	&=\beta\sum_{\vec{k},\vec{k}'}\dfrac{n(\epsilon_{\vec{k}})-n(\epsilon_{\vec{k}'})}{\epsilon_{\vec{k}}-\epsilon_{\vec{k}'}}\left|V_{\vec{k}\vec{k}'}\right|^2.
\end{align}
We applied the Matsubara summation $\sum_n(iv_n-\epsilon_{\vec{k}})^{-1}(iv_n-\epsilon_{\vec{k}'})^{-1}=\beta\dfrac{n(\epsilon_{\vec{k}})-n(\epsilon_{\vec{k}'})}{\epsilon_{\vec{k}}-\epsilon_{\vec{k}'}}$ in the last line. Defining the effective coupling strengths as $V\equiv\dfrac{\Omega^2}{4\Delta}$ and $U\equiv\dfrac{g^2}{4\Delta}$, the second-order coefficient $\chi$ is thus in the form of
\begin{align}\label{susce}
	\chi&=\dfrac{VU}{2N}\sum_{\vec{k},\vec{k}'}\dfrac{n(\epsilon_{\vec{k}})-n(\epsilon_{\vec{k}'})}{\epsilon_{\vec{k}}-\epsilon_{\vec{k}'}}\delta_{\vec{k},\vec{k}'\pm\vec{k}_{c}}\notag\\
	&=\dfrac{VU}{N}\sum_{\vec{k},\vec{k}'}\dfrac{n(\epsilon_{\vec{k}})}{\epsilon_{\vec{k}}-\epsilon_{\vec{k}'}}\delta_{\vec{k},\vec{k}'\pm\vec{k}_{c}}\notag\\
	&=\dfrac{VU}{E_rN}(2\sum_{\vec{k}}\dfrac{n(\epsilon_{\vec{k}})}{\vec{k}^2-(\vec{k}+\vec{k}_c)^2}).
\end{align}
In the last equation, we replace the wave vector $\vec{k}$ and $\vec{k}_c$ with $k_c\vec{k}$ and $k_c\vec{k}_c$, respectively, making the redefined wave vector $\vec{k}$ dimensionless and the redefined $\vec{k}_c$ as a unit vector. The same result has been reported in Ref.~\cite{PhysRevLett.112.143002,PhysRevLett.112.143003,PhysRevLett.112.143004}. 

Next, we will continue to calculate the fourth-order contribution, through evaluating the following Matsubara summations for preparation:
\begin{align}
	&\beta^{-1}\sum_n(iv_n-x_1)^{-2}(iv_n-x_2)^{-2}\notag\\
	=&-\dfrac{\beta n(x_1)[1-n(x_1)]}{(x_1-x_2)^2}-\dfrac{2n(x_1)}{(x_1-x_2)^3}-\dfrac{\beta n(x_2)[1-n(x_2)]}{(x_2-x_1)^2}-\dfrac{2n(x_2)}{(x_2-x_1)^3}\rightarrow-\dfrac{2n(x_1)}{(x_1-x_2)^3}-\dfrac{2n(x_2)}{(x_2-x_1)^3},
\end{align}
\begin{align}
	&\beta^{-1}\sum_n(iv_n-x_1)^{-2}(iv_n-x_2)^{-1}(iv_n-x_3)^{-1}\notag\\
	=&\dfrac{n(x_2)}{(x_2-x_1)^2(x_2-x_3)}+\dfrac{n(x_3)}{(x_3-x_1)^2(x_3-x_2)}-\dfrac{\beta n(x_1)[1-n(x_1)]}{(x_1-x_2)(x_1-x_3)}-\dfrac{n(x_1)}{(x_1-x_2)^2(x_1-x_3)}-\dfrac{n(x_1)}{(x_1-x_2)(x_1-x_3)^2}\notag\\
	\rightarrow&\dfrac{n(x_2)}{(x_2-x_1)^2(x_2-x_3)}+\dfrac{n(x_3)}{(x_3-x_1)^2(x_3-x_2)}-\dfrac{n(x_1)}{(x_1-x_2)^2(x_1-x_3)}-\dfrac{n(x_1)}{(x_1-x_2)(x_1-x_3)^2}.
\end{align}
The signal ``$\approx$'' means we ignore the terms like $\sim\beta n(x_1)[1-n(x_1)]$, recovering the fourth order pertubative theory in the zero-temperature limit. Beacuse such terms will be exactly cancelled by other contributions originated from the dependence between the chemical potential and the order parameter as $\dfrac{\partial\mu}{\partial\psi^2}$, which will be explicitly drivated in later section on finite temperature case. Then the coefficient $\eta$ can be divided into three different parts as
\begin{align}
	&\eta=\dfrac{1}{4}\times2\times\dfrac{U^2V^2}{E_r^3N^2}(\tilde{\eta}_1+\tilde{\eta}_2+\tilde{\eta}_3),
\end{align}
with each part taking the form of
\begin{align}
	\tilde{\eta}_1&=\beta^{-1}E_r^3\sum_{\vec{k},n}(iv_n-\epsilon_{\vec{k}})^{-1}(iv_n-\epsilon_{\vec{k}+\vec{k}_c})^{-1}(iv_n-\epsilon_{\vec{k}+2\vec{k}_c})^{-1}(iv_n-\epsilon_{\vec{k}+\vec{k}_c})^{-1}\notag\\
	&=\sum_{\vec{k}}\dfrac{n(\epsilon_{\vec{k}+\vec{k}_c})}{[(\vec{k}+\vec{k}_c)^2-\vec{k}^2]^2[(\vec{k}+\vec{k}_c)^2-(\vec{k}-\vec{k}_c)^2]}+\sum_{\vec{k}}\dfrac{n(\epsilon_{\vec{k}-\vec{k}_c})}{[(\vec{k}-\vec{k}_c)^2-\vec{k}^2]^2[(\vec{k}-\vec{k}_c)^2-(\vec{k}+\vec{k}_c)^2]}\notag\\
	&-\sum_{\vec{k}}\dfrac{n(\epsilon_{\vec{k}})}{[\vec{k}^2-(\vec{k}+\vec{k}_c)^2]^2[\vec{k}^2-(\vec{k}-\vec{k}_c)^2]}-\sum_{\vec{k}}\dfrac{n(\epsilon_{\vec{k}})}{[\vec{k}^2-(\vec{k}+\vec{k}_c)^2][\vec{k}^2-(\vec{k}-\vec{k}_c)^2]^2},\label{eq:eta1}
\end{align}
\begin{align}
	&\tilde{\eta}_2=\beta^{-1}E_r^3\sum_{\vec{k},n}(iv_n-\epsilon_{\vec{k}})^{-1}(iv_n-\epsilon_{\vec{k}+\vec{k}_c})^{-1}(iv_n-\epsilon_{\vec{k}})^{-1}(iv_n-\epsilon_{\vec{k}-\vec{k}_c})^{-1}=\tilde{\eta}_1,\label{eq:eta2}
\end{align}
\begin{align}
	&\tilde{\eta}_3=\beta^{-1}E_r^3\sum_{\vec{k},n}(iv_n-\epsilon_{\vec{k}})^{-2}(iv_n-\epsilon_{\vec{k}+\vec{k}_c})^{-2}=-\sum_{\vec{k}}\dfrac{4n(\epsilon_{\vec{k}})}{[\vec{k}^2-(\vec{k}+\vec{k}_c)^2]^3}.
\end{align}

These three different parts represent three different kinds of fermionic momentum transfer in two-photon processes, which are shown in~\ref{eta_schematic}(a-c) respectively. We can easily see that $\tilde{\eta}_1$ and $\tilde{\eta}_2$ are two-photon scattering processes, while $\tilde{\eta}_3$ is the repeated contribution of single-photon scattering. In the last equation above, we can easily check that $\tilde{\eta}_1=\tilde{\eta}_2$ by changing $\vec{k}+\vec{k}_c\rightarrow\vec{k}$. Next, we need to calculate the summation of $\vec{k}$ above. We first focus on 1D systems and change the summation into a 1D integral, giving an extra factor $\dfrac{L}{2\pi}=\dfrac{N}{2k_F}$ ahead. Combining the following integral results,
\begin{align}
	&\sum_{\vec{k}}\dfrac{n(\epsilon_{\vec{k}+\vec{k}_c})}{[(\vec{k}+\vec{k}_c)^2-\vec{k}^2]^2[(\vec{k}+\vec{k}_c)^2-(\vec{k}-\vec{k}_c)^2]}=\sum_{\vec{k}}\dfrac{n(\epsilon_{\vec{k}-\vec{k}_c})}{[(\vec{k}-\vec{k}_c)^2-\vec{k}^2]^2[(\vec{k}-\vec{k}_c)^2-(\vec{k}+\vec{k}_c)^2]}\notag\\
	=&\dfrac{N}{2k_F}\int_{-k_F}^{k_F}\dfrac{dk}{[k^2-(k-1)^2]^2[k^2-(k-2)^2]}=\dfrac{N}{2k_F}\left[\dfrac{1}{8}\dfrac{2k_F}{k_F^2-1/4}+\dfrac{1}{4}(ln\left|\dfrac{k_F+1/2}{-k_F+1/2}\right|-ln\left|\dfrac{k_F+1}{-k_F+1}\right|)\right],\\
	&\sum_{\vec{k}}\dfrac{n(\epsilon_{\vec{k}})}{[\vec{k}^2-(\vec{k}+\vec{k}_c)^2]^2[\vec{k}^2-(\vec{k}-\vec{k}_c)^2]}=\sum_{\vec{k}}\dfrac{n(\epsilon_{\vec{k}})}{[\vec{k}^2-(\vec{k}+\vec{k}_c)^2][\vec{k}^2-(\vec{k}-\vec{k}_c)^2]^2}\notag\\
	=&\dfrac{N}{2k_F}\int_{-k_F}^{k_F}\dfrac{dk}{[k^2-(k+1)^2]^2[k^2-(k-1)^2]}=\dfrac{N}{2k_F}\left[\dfrac{1}{8}\dfrac{2k_F}{k_F^2-1/4}+\dfrac{1}{4}ln\left|\dfrac{k_F-1/2}{-k_F-1/2}\right|\right],\\
	&\sum_{\vec{k}}\dfrac{n(\epsilon_{\vec{k}})}{[\vec{k}^2-(\vec{k}+\vec{k}_c)^2]^3}=\dfrac{N}{2k_F}\int_{-k_F}^{k_F}\dfrac{dk}{[k^2-(k+1)^2]^3}=-\dfrac{N}{2k_F}\dfrac{1}{16}\dfrac{2k_F}{(1/4-k_F^2)^2},
\end{align}
the coefficients $\chi$ and $\eta$ can be obtained as
\begin{align}
	\eta&=\dfrac{1}{4}\times2\times\dfrac{U^2V^2}{E_r^3N^2}\dfrac{N}{2k_F}\left(2\ln\left|\dfrac{k_F+1/2}{-k_F+1/2}\right|-\ln\left|\dfrac{k_F+1}{-k_F+1}\right|+\dfrac{k_F}{2(1/4-k_F^2)^2}\right),\\
	\chi&=\dfrac{VU}{E_rN}\dfrac{N}{2k_F}\ln\left|\dfrac{-k_F+1/2}{k_F+1/2}\right|.
\end{align}

Therefore, the superradiant order parameter $\psi$ satisfies

		\begin{align}\label{order_para}
			A\dfrac{\psi^2}{N}=-\dfrac{k_F+2(B/A)\ln\left|\dfrac{-k_F+1/2}{k_F+1/2}\right|}{8(B/A)^2\left(2\ln\left|\dfrac{k_F+1/2}{-k_F+1/2}\right|-\ln\left|\dfrac{k_F+1}{-k_F+1}\right|+\dfrac{k_F}{2(1/4-k_F^2)^2}\right)},
		\end{align}
	
	where $A\equiv\tilde{\omega}/E_r$ and $B\equiv UV/E_r^2$ are dimensionless parameters with effective coupling strengths $V\equiv{\Omega^2}/({4\Delta})$, $U\equiv{g^2}/({4\Delta})$. It turns out that the minimum solution of free energy is located around $\psi^2\sim N/A$, indicating a macroscopic number of photons in the cavity. Meanwhile, the larger $A=\tilde{\omega}/E_r$ corresponds to less photon number, as it will cost more fermion scattering processes to excite one single photon. Notice that the analytical order parameter $\psi$ is obtained under the fourth-order perturbation theory. Thus, the right-hand side of Eq.~(\ref{order_para}) may not be valid far from the critical point, where the perturbative theory becomes less accurate.
	
	\begin{figure}\label{pd_order_para}
		\centering
		\includegraphics[width=0.9\textwidth]{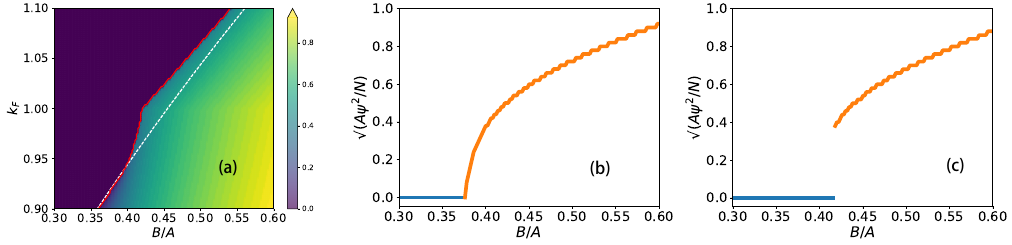}
		\hspace{-0.3cm}
		\caption{The order parameter changes with $B/A$ at fixed Fermi radius $k_F=0.918$(b) and $k_F=0.982$(c). The blue and orange parts represent the NP and SRP regions, where the order parameter is zero and nonzero, respectively.}
	\end{figure}
	
	
	When it comes to the 2D case, the fourth-order contribution can be evaluated as
	\begin{align}
		\tilde{\eta}_1^{2D}+\tilde{\eta}_2^{2D}\propto&\sum_{\vec{k}}\dfrac{n(\epsilon_{\vec{k}+\vec{k}_c})}{[(\vec{k}+\vec{k}_c)^2-\vec{k}^2]^2[(\vec{k}+\vec{k}_c)^2-(\vec{k}-\vec{k}_c)^2]}-\sum_{\vec{k}}\dfrac{n(\epsilon_{\vec{k}})}{[\vec{k}^2-(\vec{k}+\vec{k}_c)^2]^2[\vec{k}^2-(\vec{k}-\vec{k}_c)^2]}\notag\\
		\propto&N\int_{\mathcal{S}}(\dfrac{2}{1/2+k_x}-\dfrac{1}{k_x+1})d^2k=\int_{0}^{2\pi}d\theta(\dfrac{4}{1+2k\cos\theta}-\dfrac{1}{1+k\cos\theta})\int_{0}^{k_F}kdk\notag\\
		=&\dfrac{\pi}{2}[1-\Theta(\dfrac{1}{2}-k_F)\sqrt{1-4k_F^2}]-\dfrac{\pi}{2}[1-\Theta(1-k_F)\sqrt{1-k_F^2}]\notag\\
		=&\dfrac{\pi}{2}[\Theta(1-k_F)\sqrt{1-k_F^2}-\Theta(\dfrac{1}{2}-k_F)\sqrt{1-4k_F^2}]\ge0.
	\end{align}
	Therefore, we can readily get that 
	\begin{align}
		\eta^{2D}\propto\tilde{\eta}_1^{2D}+\tilde{\eta}_2^{2D}+\tilde{\eta}_3^{2D}\ge\tilde{\eta}_3^{2D}>0.
	\end{align}
	By the way, the susceptibility $\chi^{2D}$ is obtained by 
	\begin{align}
		\chi^{2D}=\dfrac{VU}{E_rN}(2\sum_{\vec{k}}\dfrac{n(\epsilon_{\vec{k}})}{\vec{k}^2-(\vec{k}+\vec{k}_c)^2})\propto-[1-\Theta(\dfrac{1}{2}-k_F)\sqrt{1-4k_F^2}].
	\end{align}

As we know that the intersection between two curves $\tilde{\omega}+4\chi=0$ and $\eta=0$ gives the tricritical point of the system, we find that the 1D and 2D systems will demonstrate different critical behaviors. The 2D systems won't demonstrate the tricritical phenomenon since $\eta>0$ is always valid. While $\eta$ for 1D systems will undergo infrared divergence behaviors in both $k_F\sim1/2$ and $k_F\sim1$ with opposite signs. In the region of $k_F\in(0,1/2]$, $\eta>0$ is valid. And we can easily verify that $\eta<0$ when $k_F>1$ as:
\begin{align}
	\eta\bigg|_{k_F>1}&\propto2\ln\left|\dfrac{k_F+1/2}{-k_F+1/2}\right|-\ln\left|\dfrac{k_F+1}{-k_F+1}\right|+\dfrac{k_F}{2(1/4-k_F^2)^2}\notag\\
	&=\ln(\dfrac{(x+1)^2}{(x-1)^2})-\ln(\dfrac{x+2}{x-2})+\dfrac{1}{(x-1)^2}-\dfrac{1}{(x+1)^2}\notag\\
	&=h(x^2+2x)-h(x^2-2x)<0.
\end{align}
In the second line, we have replaced $2k_F$ with a new variable $x$, and the expression is evaluated in the region $x>2$. By defining a new function $h(y)=\ln(1+\dfrac{1}{y})-\dfrac{1}{y+1}$ in the third line, we can easily check that $\dfrac{dh}{dy}\bigg|_{y>0}<0$, implying that the function $h(y)$ is monotonically decreasing and the final result is negative. It is also checked that $\eta$ is monotonically decreasing in the intermediate region of $k_F\in[1/2,1]$, indicating that there must be one and only one tricritical point in this system. It then turns out that the zero point of $\eta$ is achieved around the boundary of the 2nd Brillouin zone $k_F\approx0.9425$. As is depicted in Fig.~\ref{pd_order_para}(b)(c), the lower and upper boundaries in (a) display a 2nd- and 1st-order phase transition, featured by the discontinuity of the 1st and 0nd order differentiation of the order parameter, respectively.

\section{Critical scaling}
According to the final expression for the order parameter above, we can easily see that $\sqrt{A}\dfrac{\psi}{\sqrt{N}}\sim\left|(B-B_c)/A\right|^{1/2}$, where $B_c$ represents the critical point for the second order phase transition. While at the tricritical point $(B^t,k_{F}^t)$ with $B_c=B^t$, the fourth order coefficient is also zero, so the higher order (sixth order) coefficient will lead to different scaling rate as $\sqrt{A}\dfrac{\psi}{\sqrt{N}}\sim\left|(B-B_c)/A\right|^{1/4}$. 
Take logarithms on both sides of the scaling equation, we will have
$\ln(\dfrac{A\psi^2}{N})=2\nu\ln(\dfrac{B-B_c}{A})+C$, where $\nu$ is the scaling rate, taking the value $1/2$ and $1/4$ in two different cases, and $C$ is the constant number depending on the filling factor $k_F$.

In order to verify the scaling rate, we plot the critical behavior as Fig. 3(a) in main text. We choose two fixed Fermi momenta as $k_F=0.918$ and $k_F=k_F^t=0.9425$, investigating the normal critical behavior and the tricritical behavior, respectively. The blue and orange dots represent the normal critical and tricritical cases, respectively, which match well with the linear fitting lines with ramping rates $2\nu=1$ (green solid) and $2\nu=1/2$ (red solid).

In addition, we also derive the scaling law for the filling factor $k_F$. We can easily check differentiation of the second order coefficient, i.e., the numerator of Eq.~(11) in the main text against $k_F$, giving
\begin{align}
	&\dfrac{\partial}{\partial k_F}\left[k_F+2(B/A)\ln\left|\dfrac{-k_F+1/2}{k_F+1/2}\right|\right]=1+\dfrac{2B/A}{k_F^2-1/4}\ne0,
\end{align}
which indicates the scaling law as $\sqrt{A}\dfrac{\psi}{\sqrt{N}}\sim\left|k_F^c-k_F\right|^{1/2}$. By means of the same method, we numerically verify the scaling law of $\nu=1/2$ as Fig. 3(b) in main text.



\section{Searching for the global minimum of the Free energy in finite temperature case}
To investigate the whole system in finite-temperature cases, the system partition function is derived as $Z=\int\mathcal{D}[\psi,\left\{\zeta_{\vec{k}}\right\}]\exp(-S)$ with the imaginary-time action,
\begin{align}
	\label{action S}
	S&=\int_{0}^{\beta}d\tau\Biggl\{\psi^*(\partial_\tau+\tilde{\omega})\psi+\sum_{\vec{k}}\zeta_{\vec{k}}^*(\partial_\tau+\dfrac{\vec{k}^2}{2m}-\mu)\zeta_{\vec{k}}\notag\\
	&-\sum_{\vec{k}}\left[\dfrac{g\Omega}{4\Delta\sqrt{N}}(\psi+\psi^*)\sum_{s}\zeta_{\vec{k}}^*\zeta_{\vec{k}+s\vec{k}}\right]\Biggr\}.
\end{align}
Here, the complex scalar field $\psi$ refers to the bosonic mode $a$, and the Grassmann number $\zeta_{\vec{k}}$ corresponds to the fermionic modes $c_{\vec{k}}$ in the functional integral.

The path integral for the fermions can be treated first by assuming that the bosonic field is independent of $\tau$, i.e., the fermion part is separable with the boson part wave function. Defining $S\equiv S_0+S_1$, where $S_0$ is the action for noninteractive boson modes, expressed as $S_0\equiv\int_{0}^{\beta}d\tau\psi^*(\partial_\tau+\tilde{\omega})\psi$, the grant partition function of the fermion part becomes
\begin{align}
	\label{partition function}
	Z_G&=\int\mathcal{D}[\left\{\zeta_{\vec{k}}\right\}]\exp(-S_1)\equiv \exp(-\tilde{S}_1).
\end{align}
In statistic physics, the grant potential can be defined as
\begin{align}
	J\equiv U-TS-\bar{N}\mu=-k_BT\ln Z_G,
\end{align}
where $U$, $S$ are the internal energy and entropy of the system, respectively, $k_B$ is the Boltzmann constant which we set as $1$ in the text, and $\beta=1/(k_B T)$. The grand partition function $Z_G$ reads as
\begin{align}\label{grant partition function}
	Z_G=\prod_l(1\pm e^{-\alpha-\beta\epsilon_l})^{\pm\omega_l}.
\end{align}
Here, $\omega_l$ stands for the number of degeneracy, and $\pm$ refers to the fermion (boson) situation respectively. Such expression for $Z_G$ can also be derived from the Eq.~(\ref{partition function}) by calculate the Gaussian functional. Therefore, we can easily see that $\tilde{S}_1=-\ln(Z_G)$ is nothing but the counterpart of the grant potential in field theory.

Thus, the free energy of the system can be derived as
\begin{align}
	\label{free energy}
	\hspace{-0.2cm}
	F&\equiv U-TS=J+N\mu=\tilde{\omega}\left|\psi\right|^2-k_BT\ln Z_G+N\mu\notag\\
	&=\tilde{\omega}\left|\psi\right|^2-\beta^{-1}\sum_{\vec{k}}\ln(1+e^{-\beta(\epsilon_{\vec{k}}-\mu)})+N\mu,
\end{align}
whose minimum decides the status of the system. The Landau coefficient can be obtained by the derivative of the free energy against the optical field $\psi$ with a mean-field approximation.

\section{Iterative calculating the self-consistent equation in zero temperature case}

\subsection{Closed system}
In zero temperature limit, the Free energy is reduced to the ground state energy of the system, where we can solve the self-consistent equation through iterative method.

Different from the perturbation method which is invalid in the region away from the critical points, the self-consistent equation gives the exact solution of the order parameter $\psi$, derivated from $\dfrac{\partial \left<H_{tot}(\psi)\right>}{\partial\psi}=0$, where
\begin{align}\label{SCE}
	\psi=\dfrac{\sqrt{B}}{A\sqrt{N}}\left<\sum_{\vec{k},s}c_{\vec{k}}^\dagger c_{\vec{k}+s\vec{k}_c}\right>\equiv\dfrac{\sqrt{B}}{A\sqrt{N}}\Theta.
\end{align}
Here, $\psi\equiv\left<a\right>$, and $\Theta\equiv\left<\sum_{\vec{k},s}c_{\vec{k}}^\dagger c_{\vec{k}+s\vec{k}_c}\right>$ is known as the density wave order dependent on the $\psi$, for the reason of which it's called self-consistent equation. Generally, the density wave order $\Theta\sim N$, thus $\psi\sim\sqrt{N}$.

\subsection{Dissipative system}

The analysis above are based on the ground state of closed system. While for a dissipative cavity, we will focus on the operator dynamics of master equation, expressed as $\dfrac{d\mathcal{O}}{dt}=i[H,\mathcal{O}]+\mathcal{L}[\mathcal{O}]$. And the Lindblad superoperator $\mathcal{L}$ can be expressed as $\mathcal{L}[\mathcal{O}]=\kappa(2a^{\dagger}\mathcal{O}a-\mathcal{O}a^{\dagger}a-a^{\dagger}a\mathcal{O})$. Hence the dynamics of the optical operator is given by
\begin{align}
	\dfrac{da}{dt}=-i\tilde{\omega}a-\kappa a+i\dfrac{\sqrt{UV}}{\sqrt{N}}\sum_{\vec{k},s}c_{\vec{k}}^\dagger c_{\vec{k}+s\vec{k}_c}.
\end{align}

By calculating the expectation value on both sides, we can get the self-consistent equation for stable state as
\begin{align}\label{dissipative}
	\psi=\dfrac{\sqrt{UV}\Theta}{\sqrt{N}(\tilde{\omega}-i\kappa)}.
\end{align}
And the fermionic part will be stablized into the eigenstates of the fermionic Hamiltonian after mean-field approximation, expressed as
\begin{align}
	H_A=\sum_{\vec{k}}\dfrac{\vec{k}^2}{2m}c_{\vec{k}}^\dagger c_{\vec{k}}-\dfrac{g\Omega}{4\Delta\sqrt{N}}(\psi_s+\psi^*_s)\sum_{\vec{k},s}c_{\vec{k}}^\dagger c_{\vec{k}+s\vec{k}_c},
\end{align}
where $\psi_s$ is the solution of the self-consistent equation~(\ref{dissipative}).

It is obvious that the real part of the self-consistent equation satisfies $\text{Re}\psi={\sqrt{UV}\Theta}/{[\sqrt{N}(\tilde{\omega}+\kappa^2/\tilde{\omega})]}$, which corresponds to the counterpart in closed systems of the main text by replacing $\tilde{\omega}$ and $\psi$ with $\tilde{\omega}+\kappa^2/\tilde{\omega}$ and $\text{Re}\psi$. And the imaginary part will write as $\text{Im}\psi={\kappa}\text{Re}\psi/{\tilde{\omega}}$.

\subsection{Iterative method and multistability}
The self consistent equation~(\ref{SCE}) in closed system and ~(\ref{dissipative}) in dissipative system share the similar form, both of which can be solved with the same method.

Notice that the density wave order $\Theta$ in self-consistent equation is dependent on the order parameter $\psi$. Thus, the numerical solutions of the order parameter can be obtained through the iterative method. This means that we can initially select a test value of $\psi$ and then calculate the density wave order $\Theta$ to update $\psi$ until a convergent result is acquired. However, in some regions of the phase diagram, the order parameter is likely to converge to different values, implying that the system supports different stable states. Therefore, instead of finding the global minimum in the closed system, we are supposed to explore all the local minimums of the free energy.

\begin{figure*}[tb]
	\centering
	\includegraphics[width=1.0\textwidth]{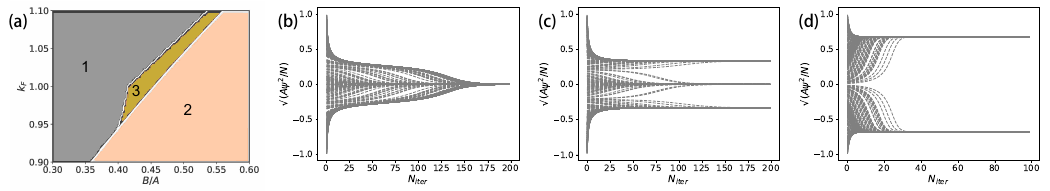}
	\caption{(a) The stable-state phase diagram in the limit $\kappa\to 0$. The grey, orange and brown regions support one, two and three stable states, respectively, which is labeled by the digits in corresponding regions. We fix the Fermi radius $k_F=0.982$, and demonstrate the results of the stable order parameter with $B/A=0.41$ in grey region 1 (b), $B/A=0.414$ in brown region 3 (c) and $B/A=0.5$ in orange region 2 (d) with the iterative method.}\label{multistable}
\end{figure*}

To facilitate comparison, we apply the numerical iterative method in the limit of $\kappa\to0$, which will not affect the qualitative results. The stable-state phase diagram is illustrated in Fig.~\ref{multistable}(a), where the digits indicate the number of stable states supported in the corresponding regions. Then we choose typical points in each region and iteratively evolve $400$ arbitrary initial test values of $\psi$ to obtain the corresponding stable order parameters, as shown in Fig.~\ref{multistable}(b-d). Through sufficient iterations, the final order parameter $\psi$ will stabilize into several branches. In Fig.~\ref{multistable}(b), we choose $B/A=0.41$ and only one stable order parameter of $\psi=0$ is obtained, indicating the NP. While for $B/A=0.414$ in Fig.~\ref{multistable}(c), although it is still in NP as shown in Fig. 2(b) of the main text, the system supports three different stable states composed of one normal state and two superradiant states. In Fig.~\ref{multistable}(d), where the dimensionless pumping strength $B/A$ is large enough, the normal states become unstable and the system enters the SRP. 

According to the analysis above, there exists a special multistable region located between the monostable region and the bistable region, which supports three different stable states. The boundary between the bistable and multistable regions coincides with the white dashed line in Fig.~\ref{pd}(b) of the main text, while the boundary between the monostable and multistable regions is slightly deviated from the SRP boundary [red solid line in Fig.~\ref{pd}(b) of the main text], which means that in dissipative cases two potential superradiant states have already been supported in NP as local stable states.

\section{Dynamic evolution of master equation}

Although the iterative method to solve the self-consistent equation is sufficient to observe the stable state both in both dissipative and non-dissipative case, but we can hardly observe the dynamic evolution by this method. Thus we combine the Heisenberg equation for the fermionic part to calculate the evolution, writing as 
\begin{align}
	&\dfrac{d\psi}{d(E_rt)}=-iA\psi-\tilde{\kappa}\psi+i\dfrac{\sqrt{B}}{\sqrt{N}}\sum_{\vec{k},s}c_{\vec{k}}^\dagger c_{\vec{k}+s\vec{k}_c},\\
	&\dfrac{dc_{\vec{k}}}{d(E_rt)}=-i\left|\vec{k}\right|^2c_{\vec{k}}+i\dfrac{\sqrt{B}}{\sqrt{N}}(\psi+\psi^*)\sum_sc_{\vec{k}+s\vec{k}_c}.
\end{align}
Here, we write the dynamics into dimensionless form and define the dimensionless dissipative rate as $\tilde{\kappa}\equiv\kappa/E_r$. $c_{\vec{k}}$ is viewed as constant number satisfying the normalized relation~\cite{PhysRevLett.131.243401}. By numerical simulations, we can verify the multistable phenomena and observe the dynamic process, where the obtained hysteresis-type evolution is displayed as Fig.~\ref{pd}(d) in the main text.

\section{Finite temperature case}

For a more general finite-temperature case, we can write the free energy of the system as 
\begin{align}\label{free_energy_finite_T}
	F=&-\beta^{-1}Tr[\ln(\beta G_0^{-1})]+N\mu+\tilde{\omega}\left|\psi\right|^2+\chi(\mu)(\psi+\psi^*)^2+\eta(\mu)(\psi+\psi^*)^4+\dots
\end{align}
Notice that in non-dissipative cavity, the expectation of optical field is real, i.e., $\psi^*=\psi$. Distinguished from the zero-temperature cases, the contribution of the chemical potential $\mu$ may make a difference at finite temperatures. Considering that $\mu$ is a function of $\psi^2$, the second-order coefficient can be calculated by
\begin{align}
	\dfrac{\partial F}{\partial(\psi^2)}=\dfrac{\partial F}{\partial(\psi^2)}\Bigg|_{\mu,\psi=0}+\dfrac{\partial F}{\partial\mu}\dfrac{\partial\mu}{\partial(\psi^2)}\Bigg|_{\psi=0}.
\end{align}

The first term is calculated with fixed chemical potential $\mu$, the result of which is $\tilde{\omega}+4\chi(\mu)$, where the expression of $\chi$ is the same as Eq.~(\ref{susce}). And the second term is an extra contribution in finite-temperature cases. Notice that we only care about the final result with order parameter $\psi=0$, So the terms $\chi(\mu)(\psi+\psi^*)^2+\eta(\mu)(\psi+\psi^*)^4+\dots$ always give zero contribution eventually. Thus, only the first two terms of Eq.~(\ref{free_energy_finite_T}) need to be calculated, which leads to
\begin{align}
	\dfrac{\partial F}{\partial\mu}\dfrac{\partial\mu}{\partial(\psi^2)}\Bigg|_{\psi=0}&=\left\{N-\beta^{-1}\dfrac{\partial}{\partial\mu}\sum_{\vec{k}}\ln[1+e^{-\beta(\epsilon_{\vec{k}}^0-\mu)}]\right\}\dfrac{\partial\mu}{\partial(\psi^2)}\Bigg|_{\psi=0}\notag\\
	&=\left\{N-\sum_{\vec{k}}[e^{\beta(\epsilon_{\vec{k}}^0-\mu^0)}+1]^{-1}\right\}\dfrac{\partial\mu}{\partial(\psi^2)}\Bigg|_{\psi=0}=0.\label{eq:FEFT}
\end{align}
In the first equation, we have applied the Matsubara summation of
\begin{align}
	\sum_{i\nu_n}\ln(\beta(-i\nu_n+\xi))=\ln(1+e^{-\beta\xi}),
\end{align}
and thus $\text{Tr}[\ln(\beta G_0^{-1})]=\sum_{\vec{k}}\ln[1+e^{-\beta(\epsilon_{\vec{k}}^0-\mu)}]$. Here, the subscript $0$ represents the value with the order parameter $\psi=0$. Moreover, we use the relation $\sum_{\vec{k}}[e^{\beta(\epsilon_{\vec{k}}^0-\mu^0)}+1]^{-1}=N$ in the second line of Eq.~(\ref{eq:FEFT}). Therefore, the expression $\chi(T,k_F)\equiv\dfrac{\partial F}{\partial(\psi^2)}$ at finite temperature $T$ makes no difference to the one in Eq.~(\ref{susce}). And we can easily check that $\chi<0$ for arbitrary temperature $T$, implying that superradiance can potentially be observed with sufficiently strong atom-cavity coupling and pumping laser intensity, even at high temperatures. 

\begin{figure}[tb]
	\centering
	\includegraphics[width=0.95\textwidth]{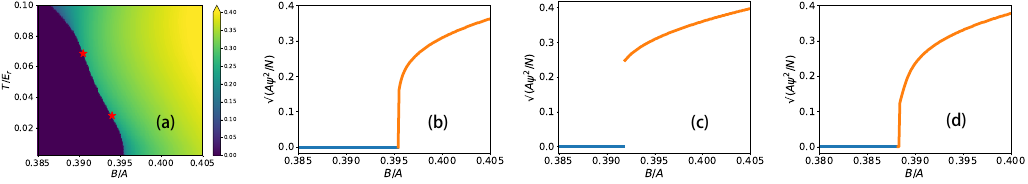}
	\caption{(a) The finite temperature phase diagram with fixed filling factor $k_F=0.938$. The order parameter changes with $B/A$ at fixed temperature $T=0.005$ (b), $T=0.05$ (c) and $T=0.09$ (d). The blue and orange parts represent the NP and SRP regions, where the order parameter is zero and nonzero, respectively.}
	\label{finiteT_pd}
\end{figure}

As demonstrated in Fig.~\ref{finiteT_pd}(a), there exist two tricritical points in the phase diagram. To make it more clear, we provide detailed order parameter plots in Fig.~\ref{finiteT_pd}(b-d). It is obvious that the typical 2nd, 1st, and 2nd order phase transitions can be observed in Fig.~\ref{finiteT_pd}(b-d), respectively, confirming the existence of two tricritical points with such system parameters at finite temperature. 


	Next, we will discuss the calculation on the fourth order coefficient $\eta'(T,k_F)\equiv\dfrac{1}{2}\dfrac{\partial^2F}{\partial(\psi^2)^2}\Bigg|_{\psi=0}$. Here we distinguish the real fourth order coefficient $\eta'$ with $\eta$ in Eq.~(\ref{free_energy_finite_T}), where the later one is obtained by assuming chemical potential $\mu$ is unchanged with $\psi^2$. Therefore, we evaluate the terms in Eq.~({\ref{free_energy_finite_T}}) one by one.
	\begin{align}
		&\dfrac{\partial^2}{\partial(\psi^2)^2}\beta^{-1}Tr[\ln(\beta G_0^{-1})]\Bigg|_{\psi=0}=\dfrac{\partial^2}{\partial(\psi^2)^2}\beta^{-1}\sum_{\vec{k}}\ln[1+e^{-\beta(\epsilon_{\vec{k}}^0-\mu)}]\notag\\
		=&\dfrac{\partial}{\partial\psi^2}\left[\sum_{\vec{k}}[e^{\beta(\epsilon_{\vec{k}}^0-\mu)}+1]^{-1}\dfrac{\partial\mu}{\partial(\psi^2)}\right]\Bigg|_{\psi=0}
		=N\dfrac{\partial^2\mu}{\partial(\psi^2)^2}\Bigg|_{\psi=0}+\sum_{\vec{k}}\beta n(\epsilon_{\vec{k}}^0)[1-n(\epsilon_{\vec{k}}^0)][\dfrac{\partial\mu}{\partial(\psi^2)}]^2\Bigg|_{\psi=0},\\
		&\dfrac{\partial^2}{\partial(\psi^2)^2}(N\mu)\Bigg|_{\psi=0}=N\dfrac{\partial^2\mu}{\partial(\psi^2)^2}\Bigg|_{\psi=0},\\
		&\dfrac{\partial^2}{\partial(\psi^2)^2}(\tilde{\omega}\left|\psi\right|^2)\Bigg|_{\psi=0}=0,\\
		&\dfrac{\partial^2}{\partial(\psi^2)^2}[\chi(\mu)(\psi+\psi^*)^2]\Bigg|_{\psi=0}=8\dfrac{\partial\chi}{\partial\mu}\dfrac{\partial\mu}{\partial(\psi^2)}\Bigg|_{\psi=0},\\
		&\dfrac{\partial^2}{\partial(\psi^2)^2}[\eta(\mu)(\psi+\psi^*)^4+\dots]=32\eta.
	\end{align}
	
	Combine the above relations, we get the analytical relation for the 
	\begin{align}\label{4th_coefficient_finit_T}
		2\eta'=\dfrac{\partial^2F}{\partial(\psi^2)^2}\Bigg|_{\psi=0}=32\eta+8\dfrac{\partial\chi}{\partial\mu}\dfrac{\partial\mu}{\partial(\psi^2)}\Bigg|_{\psi=0}-\sum_{\vec{k}}\beta n(\epsilon_{\vec{k}}^0)[1-n(\epsilon_{\vec{k}}^0)][\dfrac{\partial\mu}{\partial(\psi^2)}]^2\Bigg|_{\psi=0}.
	\end{align} 
	Till now, although we have the analytical expression for the real fourth-order coefficient $\eta'$, there still exists questions on the last two terms on the r.h.s. of Eq.~(\ref{4th_coefficient_finit_T}). In order to calculate the term $\dfrac{\partial\mu}{\partial(\psi^2)}$, we evaluate the differential against order parameter $\psi^2$ of the relation $N=\sum_{\vec{k}}[e^{\beta(\epsilon_{\vec{k}}-\mu)}+1]^{-1}$, we obtain that $0=\sum_{\vec{k}}\beta n(\epsilon_{\vec{k}})[1-n(\epsilon_{\vec{k}})][\dfrac{\partial\epsilon_{\vec{k}}}{\partial(\psi^2)}-\dfrac{\partial\mu}{\partial(\psi^2)}]d(\psi^2)$. Therefore, we can get
	\begin{align}\label{mu_against_psi}
		\dfrac{\partial\mu}{\partial(\psi^2)}=\dfrac{E_r}{\sum_{\vec{k}}n(\epsilon_{\vec{k}})[1-n(\epsilon_{\vec{k}})]}\sum_{\vec{k}}n(\epsilon_{\vec{k}})[1-n(\epsilon_{\vec{k}})]\dfrac{\partial\epsilon_{\vec{k}}}{\partial(\psi^2)}.
	\end{align}
	By means of the second order perturbation theory, we can get the differential of kinetic energy as $\dfrac{\partial\epsilon_{\vec{k}}}{\partial(\psi^2)}=4\dfrac{B}{N}[\dfrac{1}{k^2-(k+1)^2}+\dfrac{1}{k^2-(k-1)^2}]$, where we recall that $B\equiv UV/E_r^2$.
	
	Next we can easily express the differential of $\chi$ as
	\begin{align}
		\dfrac{\partial\chi}{\partial\mu}=\dfrac{VU}{E_rN}2\beta\sum_{\vec{k}}n(\epsilon_{\vec{k}}^0)[1-n(\epsilon_{\vec{k}}^0)]\dfrac{1}{k^2-(k+1)^2}.
	\end{align}
	Thus we can give the full expression of the last two terms in Eq.~\ref{4th_coefficient_finit_T} as
	\begin{align}
		8\dfrac{\partial\chi}{\partial\mu}\dfrac{\partial\mu}{\partial(\psi^2)}\Bigg|_{\psi=0}=\dfrac{32B^2E_r^2\beta}{N^2\sum_{\vec{k}}n(\epsilon_{\vec{k}}^0)[1-n(\epsilon_{\vec{k}}^0)]}\left\{\sum_{\vec{k}}n(\epsilon_{\vec{k}}^0)[1-n(\epsilon_{\vec{k}}^0)]\dfrac{2}{k^2-(k+1)^2}\right\}^2,\\
		\sum_{\vec{k}}\beta n(\epsilon_{\vec{k}}^0)[1-n(\epsilon_{\vec{k}}^0)][\dfrac{\partial\mu}{\partial(\psi^2)}]^2\Bigg|_{\psi=0}=\dfrac{16B^2E_r^2\beta}{N^2\sum_{\vec{k}}n(\epsilon_{\vec{k}}^0)[1-n(\epsilon_{\vec{k}}^0)]}\left\{\sum_{\vec{k}}n(\epsilon_{\vec{k}}^0)[1-n(\epsilon_{\vec{k}}^0)]\dfrac{2}{k^2-(k+1)^2}\right\}^2.
	\end{align}
	
	Then, we  rewrite the whole expression of $\eta$ as
	\begin{align}
		\eta=\dfrac{U^2V^2}{2E_r^3N^2}\times&\Biggl\{\sum_{\vec{k}}\dfrac{4n(\epsilon_{\vec{k}+\vec{k}_c})}{[(\vec{k}+\vec{k}_c)^2-\vec{k}^2]^2[(\vec{k}+\vec{k}_c)^2-(\vec{k}-\vec{k}_c)^2]}-\sum_{\vec{k}}\dfrac{4n(\epsilon_{\vec{k}})}{[\vec{k}^2-(\vec{k}+\vec{k}_c)^2]^2[\vec{k}^2-(\vec{k}-\vec{k}_c)^2]}\notag\\
		&-\sum_{\vec{k}}\dfrac{4n(\epsilon_{\vec{k}})}{[\vec{k}^2-(\vec{k}+\vec{k}_c)^2]^3}-2\beta E_r\sum_{\vec{k}}n(\epsilon_{\vec{k}}^0)[1-n(\epsilon_{\vec{k}}^0)]\dfrac{1}{[\vec{k}^2-(\vec{k}-\vec{k}_c)^2][\vec{k}^2-(\vec{k}+\vec{k}_c)^2]}\notag\\
		&-2\beta E_r\sum_{\vec{k}}n(\epsilon_{\vec{k}}^0)[1-n(\epsilon_{\vec{k}}^0)]\dfrac{1}{[\vec{k}^2-(\vec{k}+\vec{k}_c)^2]^2}\Biggr\},
	\end{align}
	which contained the terms ignored in the previous section. Combining the above three terms, we can get the $\eta'$ in finite-temperature case.
	
	In order to evaluate the the zero temperature limit $T\rightarrow0$, we can compare the ignored two terms of $32\eta$ in section SI, writing as
	\begin{align}\label{extra_term}
		&16\times\left(-2\beta\sum_{\vec{k}}n(\epsilon_{\vec{k}}^0)[1-n(\epsilon_{\vec{k}}^0)]\dfrac{1}{[k^2-(k-1)^2][k^2-(k+1)^2]}-2\beta\sum_{\vec{k}}n(\epsilon_{\vec{k}}^0)[1-n(\epsilon_{\vec{k}}^0)]\dfrac{1}{[k^2-(k+1)^2]^2}\right)\notag\\
		=&16\times\left(-2\beta\sum_{\vec{k}}n(\epsilon_{\vec{k}}^0)[1-n(\epsilon_{\vec{k}}^0)][\dfrac{1}{(2k-1)(-2k-1)}+\dfrac{1}{(2k+1)^2}]\right),
	\end{align}
	and the contribution of the last two terms in Eq.~(\ref{4th_coefficient_finit_T}) as 
	\begin{align}\label{last_two_term}
		\dfrac{16\beta}{\sum_{\vec{k}}n(\epsilon_{\vec{k}}^0)[1-n(\epsilon_{\vec{k}}^0)]}\left\{\sum_{\vec{k}}n(\epsilon_{\vec{k}}^0)[1-n(\epsilon_{\vec{k}}^0)]\dfrac{2}{k^2-(k+1)^2}\right\}^2.
	\end{align}
	Here, we neglect the the complex coefficient before the expressions, and concentrate on 1D gases. Next, we consider the differential 
	\begin{align}
		\dfrac{\partial}{\partial\mu}\sum_{\vec{k}}n(\epsilon_{\vec{k}}^0)=\sum_{\vec{k}}[-\dfrac{\partial}{\partial\epsilon_{\vec{k}}^0}n(\epsilon_{\vec{k}}^0)]=\sum_{\vec{k}}\beta n(\epsilon_{\vec{k}}^0)[1-n(\epsilon_{\vec{k}}^0)],
	\end{align}
	in zero temperature limit. It is easily checked that $\int_{0}^{\infty}d\epsilon_{\vec{k}}^0[-\dfrac{\partial}{\partial\epsilon_{\vec{k}}^0}n(\epsilon_{\vec{k}}^0)]=1$, and if $\epsilon_{\vec{k}}^0\neq\mu$, we have $-\dfrac{\partial}{\partial\epsilon_{\vec{k}}^0}n(\epsilon_{\vec{k}}^0)=0$. Therefore, a typical $\delta$-function is obtained as $-\dfrac{\partial}{\partial\epsilon_{\vec{k}}^0}n(\epsilon_{\vec{k}}^0)=\beta n(\epsilon_{\vec{k}}^0)[1-n(\epsilon_{\vec{k}}^0)]=\delta(\epsilon_{\vec{k}}^0-\mu)$ with $T\rightarrow0$. By means of this property, we can get that
	\begin{align}
		(\ref{extra_term})/16&=-2[\dfrac{1}{(2\sqrt{\mu}-1)(-2\sqrt{\mu}-1)}+\dfrac{1}{(2\sqrt{\mu}+1)^2}+\dfrac{1}{(-2\sqrt{\mu}-1)(2\sqrt{\mu}-1)}+\dfrac{1}{(-2\sqrt{\mu}+1)^2}]\notag\\
		&=-\dfrac{8}{(1-\sqrt{\mu})^2(1+\sqrt{\mu})^2}\\
		(\ref{last_two_term})/16&=\dfrac{1}{2}[\dfrac{2}{\sqrt{\mu}^2-(\sqrt{\mu}+1)^2}+\dfrac{2}{(-\sqrt{\mu})^2-(-\sqrt{\mu}+1)^2}]^2=\dfrac{8}{(1-\sqrt{\mu})^2(1+\sqrt{\mu})^2}.
	\end{align}
	
	Here, we ignore the constant factor $\dfrac{N}{2k_F}\dfrac{1}{\partial\epsilon_{\vec{k}}^0/\partial k}\Bigg|_{\epsilon_{\vec{k}}^0=\mu}$ for easy writing, which is generated by changing the summation into integral and the differentiation from the $\delta$-function. Therefore, we can see that the calculation in the previous section based on the 4th order perturbative theory is correct. The reason why we ignore the terms in Eq.~(\ref{extra_term}) in previous calculation, is that such terms will be exactly canceled eventually. Here we also want to emphasize that this analytical result is valid for 2D systems, where we just modify the $\sqrt{\mu}\rightarrow\sqrt{\mu}\cos(\theta)$ and consider an extra integral dimension from $\theta=0$ to $\theta=\pi$. And $\theta$ is defined as the angle between the wave vector $\vec{k}$ and $\vec{k}_c$.
	
	\section{The dependence of $\chi$ against $T$}
	For the fixed filling factor $k_F$, the susceptibility can be viewed as the function of the coupling strength and temperature $T$, which writes as $\chi=\chi(T,B;k_F)$. As we set $-\tilde{\omega}/4=\chi(T_c,B_c;k_F)$, we can get the implicit function relation between the critical temperature $T_c$ and coupling strength $B_c$. By differentiating this relation, we can get
	\begin{align}
		\dfrac{dB_c}{dT_c}=-\dfrac{\partial\chi/\partial T_c}{\partial\chi/\partial B_c}.
	\end{align}
	So, we will discuss the first order tendency of the susceptibility $\chi$ changing against $T$ in this section.
	
	Consider that the free energy of the atomic ensemble can be expressed as
	\begin{align}\label{atom_free_energy}
		F_A=-\beta^{-1}\sum_{\vec{k}}\ln[1+e^{-\beta(\epsilon_{\vec{k}}-\mu)}]+N\mu.
	\end{align}
	Notice that in this case $\epsilon_{\vec{k}}$ is the diagonalized kinetic energy depending on the order parameter $\psi^2$. Next, we need to calculate the second order differentiation $\dfrac{\partial\chi}{\partial T}\Bigg|_{\psi=0,T\rightarrow0}=\dfrac{\partial^2F_A}{\partial\psi^2\partial T}\Bigg|_{\psi=0,T\rightarrow0}$. We first notice that 
	\begin{align}
		\dfrac{\partial F_A}{\partial\psi^2}=\sum_{\vec{k}}n(\epsilon_{\vec{k}})\dfrac{\partial(\epsilon_{\vec{k}}-\mu)}{\partial\psi^2}+N\dfrac{\partial\mu}{\partial\psi^2}.
	\end{align}
	Thus, we can get 
	\begin{align}
		\dfrac{\partial^2F_A}{\partial\psi^2\partial T}\Bigg|_{\psi=0,T\rightarrow0}&=\left\{\sum_{\vec{k}}\dfrac{\partial}{\partial T}[n(\epsilon_{\vec{k}})]\dfrac{\partial(\epsilon_{\vec{k}}-\mu)}{\partial\psi^2}+\sum_{\vec{k}}n(\epsilon_{\vec{k}})\dfrac{\partial^2(\epsilon_{\vec{k}}-\mu)}{\partial\psi^2\partial T}+N\dfrac{\partial^2\mu}{\partial\psi^2\partial T}\right\}\Bigg|_{\psi=0,T\rightarrow0}\notag\\
		&=\sum_{\vec{k}}\dfrac{\partial}{\partial T}[n(\epsilon_{\vec{k}})]\dfrac{\partial(\epsilon_{\vec{k}}-\mu)}{\partial\psi^2}\Bigg|_{\psi=0,T\rightarrow0}\notag\\
		&=\sum_{\vec{k}}\beta^2n(\epsilon_{\vec{k}})[1-n(\epsilon_{\vec{k}})]\dfrac{\partial[\beta(\epsilon_{\vec{k}}-\mu)]}{\partial\beta}\dfrac{\partial(\epsilon_{\vec{k}}-\mu)}{\partial\psi^2}\Bigg|_{\psi=0,T\rightarrow0},
	\end{align}
	where the second equation we apply the property that the kinetic energy is independent on the temperature $T$, and find that the second and third terms cancel exactly.
	Considering that the total atom number is independent of the system parameter $T$ and order parameter $\psi^2$, we will have
	\begin{align}
		0&=\dfrac{\partial N}{\partial T}=\beta^2\sum_{\vec{k}}n(\epsilon_{\vec{k}})[1-n(\epsilon_{\vec{k}})]\dfrac{\partial[\beta(\epsilon_{\vec{k}}-\mu)]}{\partial\beta},\\
		0&=\dfrac{\partial N}{\partial\psi^2}=-\beta\sum_{\vec{k}}n(\epsilon_{\vec{k}})[1-n(\epsilon_{\vec{k}})]\dfrac{\partial(\epsilon_{\vec{k}}-\mu)}{\partial\psi^2}.
	\end{align}

	In aid of the $\delta$-function property of the term $\beta n(\epsilon_{\vec{k}})[1-n(\epsilon_{\vec{k}})]$, we can get $\lim_{T\rightarrow0}\beta\dfrac{\partial[\beta(\epsilon_{\vec{k}}-\mu)]}{\partial\beta}\Bigg|_{\epsilon_{\vec{k}}=\mu}=0$, and $\lim_{T\rightarrow0}\dfrac{\partial(\epsilon_{\vec{k}}-\mu)}{\partial\psi^2}\Bigg|_{\epsilon_{\vec{k}}=\mu}=0$. Thus, it becomes
	\begin{align}
		\left|\dfrac{\partial^2F_A}{\partial\psi^2\partial T}\right|\Bigg|_{\psi=0,T\rightarrow0}
		&=\sum_{\vec{k}}\beta^2n(\epsilon_{\vec{k}})[1-n(\epsilon_{\vec{k}})]\dfrac{\partial[\beta(\epsilon_{\vec{k}}-\mu)]}{\partial\beta}\dfrac{\partial(\epsilon_{\vec{k}}-\mu)}{\partial\psi^2}\Bigg|_{\psi=0,T\rightarrow0}\notag\\
		&=\beta\dfrac{\partial[\beta(\epsilon_{\vec{k}}-\mu)]}{\partial\beta}\dfrac{\partial(\epsilon_{\vec{k}}-\mu)}{\partial\psi^2}\Bigg|_{\psi=0,\epsilon_{\vec{k}}=\mu,T\rightarrow0}=0\times0=0.
	\end{align}
	Here, we have omitted the common constant factor $\dfrac{N}{2k_F}\dfrac{1}{\partial\epsilon_{\vec{k}}/\partial k}\Bigg|_{\psi=0,\epsilon_{\vec{k}}=\mu}$, which makes no difference to the conclusion.
	
	In summary, we obtain that $\dfrac{\partial\chi}{\partial T}\Bigg|_{\psi=0,T\rightarrow0}=0$, so we have $\dfrac{dB_c}{dT_c}\Bigg|_{T_c\rightarrow0}=0$. This relation tells us a nontrivial fact that the critical boundary in finite temperature case won't deviate from the zero temperature critical point, at least in terms of the first order of temperature. Thus, we can give that scaling rate $\lim_{\Delta T\rightarrow0}\Delta B_c\sim \Delta T^\nu$, and $\nu>1$, where $\Delta B_c\equiv B_c\Bigg|_{T=\Delta T}-B_c\Bigg|_{T=0}$.

\twocolumngrid

\bibliography{references.bib}

\begin{thebibliography}{49}
\bibitem [{ind()}]{independentmu}%
  \BibitemOpen
  \href@noop {} {}\bibinfo {note} {It can also be verified by inserting the
  grant partition function ~(\ref{grant partition function}) into
  Eq.~(\ref{free energy}) as \begin{align}
  F(\beta\to\infty)&=\tilde{\omega}\left|\psi\right|^2-\lim_{\beta\to\infty}\beta^{-1}\sum_{\vec{k}}\ln(1+e^{-\beta(\epsilon_{\vec{k}}-\mu)})+N\mu\notag\\
  &=\tilde{\omega}\left|\psi\right|^2+\lim_{\beta\to\infty}\sum_{\vec{k}}\dfrac{\epsilon_{\vec{k}}-\mu}{e^{\beta(\epsilon_{\vec{k}}-\mu)}+1}+N\mu\notag\\
  &=\tilde{\omega}\left|\psi\right|^2+\sum_{\left|\vec{k}\right|\le
  k_F}\epsilon_{\vec{k}}. \end{align}}\BibitemShut {NoStop}%
\end{thebibliography}
\end{document}